\documentclass[preprint]{JHEP3} 

\JHEPspecialurl{http://jhep.sissa.it/JOURNAL/JHEP3.tar.gz}

\usepackage{epsfig,multicol,mathbbol,amsmath,graphicx,mathrsfs,axodraw}

\newcommand\fverb{\setbox\pippobox=\hbox\bgroup\verb}
\newcommand\fverbdo{\egroup\medskip\noindent%
			\fbox{\unhbox\pippobox}\ }
\newcommand\fverbit{\egroup\item[\fbox{\unhbox\pippobox}]}
\newbox\pippobox
\newcommand{\nin}{\noindent}

\newcommand{\be}{\begin{equation}}
\newcommand{\ee}{\end{equation}}
\newcommand{\bea}{\begin{eqnarray}}
\newcommand{\eea}{\end{eqnarray}}
\newcommand{\bml}{\begin{mathletters} \baselineskip 10pt}
\newcommand{\eml}{\baselineskip 12pt \end{mathletters}}
\newcommand{\nn}{\nonumber}

\newcommand{\bra}{\langle}
\newcommand{\ket}{\rangle}

\newcommand{\pprime}{\prime \prime}

\newcommand{\Px}{P_{\ssvc{x}}}
\newcommand{\Py}{P_{\ssvc{y}}}
\newcommand{\Lx}{L_{\ssvc{x}}}
\newcommand{\Ly}{L_{\ssvc{y}}}
\newcommand{\Lz}{L_{\ssvc{z}}}
\newcommand{\Kx}{K_{\ssvc{x}}}

\newcommand{\Mx}{M_{\ssvc{x}}}

\newcommand{\sfrac}[2]{{\textstyle \frac{#1}{#2}}}
\newcommand{\pad}[2]{\frac{\partial #1}{\partial #2}}

\newcommand{\vc}[1]{\mbox{\boldmath$#1$}}
\newcommand{\svc}[1]{\mbox{\footnotesize\boldmath$#1$}}
\newcommand{\ssvc}[1]{\mbox{\scriptsize\boldmath$#1$}}

\newcommand{\pol}[1]{\mathfrak{P}_{\svc{#1}}}
\newcommand{\charlink}[3]{X_{#1; \, \ssvc{#2}\ssvc{#3}}}
                               
\newcommand{\tr}{\mbox{tr}}

\renewcommand{\arraystretch}{1.5}

\title{Effective lattice theories for Polyakov loops}

\author{Leander Dittmann, Thomas Heinzl\thanks{Supported by DFG.}~
		and Andreas Wipf\\
	Theoretisch--Physikalisches Institut,
        Friedrich--Schiller--Universit\"at Jena, Max--Wien--Platz 1,
        07743 Jena, Germany\\
	E-mail: \email{l.dittmann@tpi.uni-jena.de},
        \email{t.heinzl@tpi.uni-jena.de}, \email{a.wipf@tpi.uni-jena.de}}

\preprint{FSU--TPI/06/03}      

\abstract{We derive  effective actions for $SU(2)$ Polyakov loops using
inverse Monte Carlo techniques. In a first approach, we determine the
effective couplings by requiring that the effective ensemble reproduces
the single--site distribution of the Polyakov loops. The latter is flat below the
critical temperature implying that the (untraced) Polyakov loop is distributed
uniformly over its target space, the $SU(2)$ group manifold. This allows
for an \textit{analytic} determination of the Binder cumulant and the
distribution of the mean--field, which turns out to be approximately
Gaussian. In a second approach, we employ novel lattice
Schwinger--Dyson equations which reflect the $SU(2) \times SU(2)$
invariance of the functional Haar measure. Expanding the effective
action in terms of $SU(2)$ group characters makes the numerics
sufficiently stable so that we are able to extract a total number of 14
couplings. The resulting action is short--ranged and reproduces the
Yang--Mills correlators very well.}    


\begin{document} 


\section{Introduction}

The deconfinement phase transition in pure Yang--Mills theory
\cite{polyakov:78,susskind:79} is controlled by the dynamics of the
Polyakov loop variable $\pol{x}$. Above a critical temperature $T_c$,
the singlet part $\Lx \equiv \tr \, \pol{x}/2$ develops a nonvanishing
vacuum expectation value (VEV). In this high--temperature phase one expects
to find a plasma of liberated gluons (and, in QCD, also quarks). The
VEV of $\Lx$ thus represents an order parameter associated with
spontaneous symmetry breaking. The symmetry in question is a global 
$\mathbb{Z}_N$ symmetry, $\mathbb{Z}_N$ being the center of the gauge
group $SU(N)$. While the Yang--Mills action is center symmetric, $\Lx$,
although gauge invariant, transforms nontrivially, $\Lx \to z\Lx$, $z \in
\mathbb{Z}_N$. Combining renormalization group ideas and dimensional
reduction,  Svetitsky and Yaffe have conjectured  that
finite--temperature $SU(N)$   Yang--Mills theory in $d$ dimensions is in
the universality class of a $\mathbb{Z}_N$ spin model in dimension $d-1$
\cite{svetitsky:82,yaffe:82}. For some recent and rather sophisticated
confirmations of the statement on the lattice the reader is referred to 
\cite{caselle:96,gliozzi:97,pepe:02,forcrand:03}.

The universality argument implies that effective field theory methods
may be put to use. It should make sense to map the  microscopic theory,
here Yang--Mills, onto a macroscopic one, described by an effective
action with  $\mathbb{Z}_N$ symmetry. For gauge group $SU(2)$, for
instance, one can try to coarse--grain the gauge fields all the way down to
$\mathbb{Z}_2$ Ising spins \cite{polonyi:82,okawa:88,fortunato:01}.
An intermediate procedure is to establish an effective action for the
Polyakov loop variable itself \cite{banks:83,ogilvie:84,svetitsky:86}. This
may be achieved analytically using strong--coupling or, equivalently,
high--temperature expansions \cite{ogilvie:84,caselle:95,billo:96}.
Doing so for $SU(2)$, one obtains a local effective action depending on all
characters $\chi_j (\pol{x})$ \cite{caselle:95,billo:96}.  The index $j \in
\mathbb{N}/2$ labels the  irreducible representations of $SU(2)$.  In
this most elementary case, $\chi_j$ can be expressed in terms of powers
of $\Lx$ (the character of the fundamental representation, $j=1/2$). For
larger gauge groups, however, more and more characters/representations become
relevant. This fact has recently been employed for model 
building, regarding the untraced holonomy $\pol{x}$ \cite{pisarski:00}
or, equivalently, its eigenvalues \cite{meisinger:02a} as the
fundamental degrees of freedom. We parametrize the (lattice) effective
action as follows,
\be
\label{S_EFF}
  S_{\mathrm{eff}} = \sum_a \lambda_a S_a \; ,
\ee
with center--symmetric operators $S_a$ and effective couplings
$\lambda_a$ to be determined. As stated above, for $SU(2)$ it is
sufficient to work with only the traced Polyakov loop, $\Lx$. The
effective action will then have the form \cite{svetitsky:86},
\be
\label{S_SVET}
  S_{\mathrm{eff}}[\Lx] = \sum_{\ssvc{x}} V [\Lx^2] + \sum_{\ssvc{x}
  \ssvc{y}} \Lx K_{\ssvc{x}\ssvc{y}}^{(2)} \Ly +
  \sum_{\ssvc{x}\ssvc{y} \ssvc{u}\ssvc{v}} \Lx \Ly
  K_{\ssvc{x}\ssvc{y} \ssvc{u}\ssvc{v}}^{(4)}  L_{\ssvc{u}} L_{\ssvc{v}} +
  \ldots \; .
\ee
The kernels $K^{(a)}$ depend on the couplings $\lambda_a$ and the
temperature. By construction, the $\mathbb{Z}_2$ center symmetry ($\Lx
\to - \Lx$) is manifest. Note that the representation (\ref{S_SVET}) is
rather general and leaves room for a plethora of operators, the compact
continuous variable $\Lx \in [-1 , 1]$ being dimensionless. Later on, it will
therefore be crucial to choose an appropriate subset of all possible
operators in order to capture the essential physics. In this respect it
turns out useful to follow \cite{pisarski:00} and view the effective
action (\ref{S_EFF}) as being embedded into a `sigma model' depending 
on $\pol{x}$, $ S_{\mathrm{eff}} [L] \equiv  S_{\mathrm{eff}}
[\pol{x}]$. This yields an additional global $SU(2)$ symmetry, 
\be
  \pol{x} \to g \, \pol{x} \, g^{-1} \; , \quad g \in SU(2) \; ,
  \label{ADJ} 
\ee
which is a remnant of the underlying $SU(2)$ gauge invariance. The Haar
measure $\mathcal{D} \pol{x}$ has an even larger symmetry, namely $SU(2)
\times SU(2)$, corresponding to the transformation law
\be
   \pol{x} \to g \, \pol{x} \, h \; , \quad g,h \in SU(2) \; .
   \label{LR} 
\ee  
The invariance of the measure leads to novel Schwinger--Dyson identities
which will be an important ingredient in our derivation of the effective
couplings $\lambda_a$ inherent in (\ref{S_EFF}). 

The paper is organized as follows. In Section~2 we derive exact
(lattice) Schwinger--Dyson equations from the invariance of the Haar
measure $\mathcal{D} \pol{x}$. We proceed by analysing the single--site
distribution of the Polyakov loop variable $\Lx$ in Section~3. This yields
a semianalytic method to determine all couplings $\lambda_a$ apart from
the one of the hopping term, $\lambda_0$. The latter is obtained in
Section~4 using the Schwinger--Dyson equations which are also employed
to check the resulting effective action. In Section~5, we determine the
effective potential in the symmetric phase from the single--site
distribution. Finally, in Section 6, we perform an extensive numerical
analysis to improve the effective action by including a maximum number
of 14 operators. Some technicalities concerning the analysis of
histograms are relegated to an appendix.

\section{Haar measure and Schwinger--Dyson identities}

The Polyakov loop variable on the lattice is given by a holonomy
or parallel transport connecting the (periodic) boundaries in temporal
direction,                      
\be
\label{POLDEF}
  \pol{x} \equiv \prod_{t=1}^{N_t} U_{t, \ssvc{x}; 0} \; ,
\ee
where the $U$'s are the standard link variables on a lattice of size
$N_t \times N_s^3$. The effective action for the Polyakov loops is
obtained by inserting unity into the 
Yang--Mills partition function, such that (the trace of) (\ref{POLDEF})
is imposed as a constraint,
\bea
  Z_{\mathrm{YM}} &=& \int \mathcal{D}U \, \exp(-S_W [U]) \nn \\
  &=& \int \mathcal{D}U \, \mathcal{D}\pol{} \, \delta \Big( \tr \,
  \pol{x}- \tr   \prod_{t=1}^{N_t} U_{t, \ssvc{x}; 0} \Big) \, \exp(-S_W
  [U])  \nn \\
  &\equiv& \int \mathcal{D} \pol{} \exp(-S_{\mathrm{eff}}[\pol{}]) \; ,
\eea
with $\mathcal{D}U$ and $\mathcal{D}\pol{}$ the appropriate Haar
measures (see below) and $S_W$ the standard Wilson action. Of course,
the integration over link variables $U$ in the last step cannot be
performed exactly. For this reason one has to resort to effective
actions as given by (\ref{S_EFF}) and (\ref{S_SVET}), for instance
\cite{svetitsky:82,yaffe:82,svetitsky:86}. Using inverse Monte--Carlo
(IMC) techniques, it should be possible to determine a reasonable effective
action from Yang--Mills configurations.

The main ingredient for this procedure are the Schwinger--Dyson
equations associated with the symmetry of the measure
$\mathcal{D}\pol{}$ under (\ref{LR}).  To derive those we
choose the parametrization,  
\be
\label{PARA}
  \pol{x} \equiv \Px^0 \, \Eins + i \tau^a \Px^a \equiv \Px^\mu
  \sigma^\mu  \; , 
\ee
which is in $SU(2)$, $\pol{x}^\dagger \pol{x} = \Eins$, if the components
$P_{\ssvc{x}}^\mu$ define a three--sphere $S^3$ according to 
\be
  \Px^\mu \Px^\mu = (\Px^0)^2 + \Px^a \Px^a = 1 \; .
\ee
We mention in passing that the points $\vc{x}$ where the Polyakov loop
is given by center elements, $\pol{x} = \pm \Eins$, correspond to the
positions of monopoles in the Polyakov gauge
\cite{reinhardt:97b,ford:98,jahn:98}, a particular realization of
`t~Hooft's Abelian projections \cite{thooft:81a}. 

In terms of the coordinates (\ref{PARA}), the traced Polyakov loop
becomes $\Lx = \Px^0$, while the functional Haar measure can be written
as 
\be
\label{HAAR}
  \mathcal{D}\pol{} \equiv \prod_{\ssvc{x}} d^4\Px \; \delta (\Px^\mu
  \Px^\mu - 1) \; .
\ee
Obviously, this is invariant under rotations $R \in SO(4)$ generated by
the angular momenta 
\be
  \Lx^{\mu \nu} \equiv -i \left( \Px^\mu \pad{}{\Px^\nu} - \Px^\nu
  \pad{}{\Px^\mu}   \right) \; .
\ee
These can be split up into `electric' and `magnetic' components (or
`boosts' and 3$d$ `rotations'),
\bea
  i\Lx^{0a} &\equiv& \Px^0 \pad{}{\Px^a} - \Px^a \pad{}{\Px^0} \equiv
  i\Kx^a \; , \\
  i \Lx^{ab} &\equiv& \Px^a \pad{}{\Px^b} - \Px^b \pad{}{\Px^a} \equiv i
  \epsilon^{abc} \Lx^c \; , \quad \Lx^a \equiv \frac{1}{2}
  \epsilon^{abc} \Lx^{bc} \; .
\eea
Summarizing, the $SO(4)$ generators $\Lx^{\mu\nu}$ rotate the
four--vector $\Px^\mu$, while the $SO(3)$ generators $\Lx^a$ rotate the
three--vector $\Px^a$.  The self-- and anti--selfdual combinations,
\bea
  M_{\ssvc{x}}^a \equiv \frac{1}{2} (\Lx^a - \Kx^a) \; , \\
  N_{\ssvc{x}}^a \equiv \frac{1}{2} (\Lx^a + \Kx^a) \; ,
\eea
generate left and right multiplication, respectively,
\be
  \pol{x} \to g \, \pol{x} \; , \quad \pol{x} \to \pol{x} \, h \; ,
  \quad g,h \in SU(2) \; .
\ee
Global $SU(2)$ (gauge) transformations of the Polyakov loop as given by
(\ref{ADJ}) are generated by $\Lx^{ab}$ (or $\Lx^a$) which do not
differentiate with respect to the trace $\Px^0$ and thus leave any
functional of $\Px^0 = \Lx$ invariant. Typical such invariants are
\be
  \Px^0 \; , \quad \Px^a \Px^a \equiv 1 - \Px^0 \Px^0 \; , \ldots \; .
\ee
The Schwinger--Dyson equations that follow from the $SO(4)$ invariance
of the Haar measure (\ref{HAAR}) are given by
\be
  \int \mathcal{D}\pol{} \, \Lx^{\mu\nu} \big\{ F[\pol{}] \exp(-
  S_{\mathrm{eff}}[\pol{}]) \big\} = 0 \; ,
\ee
where $ F[\pol{}]$ is an arbitrary functional of $\pol{x}$. As the
effective action depends on $\pol{x}$ solely through the $SU(2)$ 
invariant $\Px^0$, $S_{\mathrm{eff}}[\pol{}] \equiv
S_{\mathrm{eff}}[P^0]$,  only the generators $\Lx^{0a} \equiv \Kx^a$
lead to nontrivial  relations which can be written as  
\be
\label{SD1}
  \bra \Kx^a F[\pol{}] - F[\pol{}] \Kx^a S_{\mathrm{eff}}[\pol{}] \ket =
  0  \; ,  
\ee 
using the expectation value notation,
\be
  \bra O \ket \equiv Z^{-1} \int \mathcal{D}\pol{} \, O[\pol{}] \exp(-
  S_{\mathrm{eff}}[\pol{}]) \; .
\ee
Because $\Kx^a$ transforms like a vector under gauge rotations, (\ref{SD1})
in general will not be gauge invariant. However, we are still free to
choose the functional $F[\pol{}]$ at our will. If we pick
\be
\label{FG}
  F_{\ssvc{x}}^a [\pol{}] \equiv \Px^a \, G[P^0] \; ,
\ee
with an arbitrary functional $ G[P^0]$, we have the action of $\Kx^a$,
\be
  \Kx^a F_{\ssvc{y}}^b = \delta^{ab} \delta_{\ssvc{x}\ssvc{y}} \Px^0 G -
  \Px^a \Py^b \, G_{\ssvc{x}}^\prime \; ,
\ee
where we have denoted $G_{\ssvc{x}}^\prime \equiv \partial G / \partial
\Px^0$. Plugging this into (\ref{SD1}), setting $\vc{x} = \vc{y}$ and
taking the trace one finds the \textit{gauge invariant} Schwinger--Dyson
equations, 
\be
\label{SD2}
  \langle 3 \Px^0 G - \Px^a \Px^a (G_{\ssvc{x}}^\prime - G S_{\mathrm{eff},
  \ssvc{x}}^\prime) \rangle = 0 \; .
\ee
The same result is obtained using $F_{\ssvc{x}}^a [\pol{}] \equiv
\Kx^a \, H[P^0]$ instead of (\ref{FG}) and identifying
$H_{\ssvc{x}}^\prime \equiv - G_{\ssvc{x}}$. Let us rewrite 
(\ref{SD2}) as a functional integral,
\be
\label{SD3}
  \int \mathcal{D} \pol{} \, \left[ 3 \Px^0 G - \Px^a \Px^a
  (G_{\ssvc{x}}^\prime -   G S_{\mathrm{eff}, \ssvc{x}}^\prime) \right]
  \exp(-S_{\mathrm{eff}}) = 0 \; ,
\ee
and parametrize $\pol{x}$ according to
\be
  \pol{x} = \exp{i \tau^a \theta_{\ssvc{x}}^a} =  \Eins \cos
  \theta_{\ssvc{x}} + i 
  \tau^a n_{\ssvc{x}}^a \sin \theta_{\ssvc{x}} \; , \quad n_{\ssvc{x}}^a
  \equiv \Px^a/(\Px^b \Px^b)^{1/2} \; .
\ee
Then, the traced Polyakov loop is $\Lx \equiv \cos \theta_{\ssvc{x}}$
while the Haar measure (\ref{HAAR}) becomes
\be
  \mathcal{D} \pol{} = \prod_{\ssvc{x}} \sin^2 \theta_{\ssvc{x}}
  \, \frac{d\theta_{\ssvc{x}} \, d^2 n_{\ssvc{x}}}{4 \pi^2} \; .
\ee
As the functional integral (\ref{SD3}) only depends on invariants we can
integrate over the directions $n$ (yielding an irrelevant volume factor)
so that we are left with an integral involving only the \textit{reduced}
Haar measure,
\be
  \mathcal{D}L \equiv  \prod_{\ssvc{x}} d\theta_{\ssvc{x}} \, \sin^2
  \theta_{\ssvc{x}}  = \prod_{\ssvc{x}} d(\cos \theta_{\ssvc{x}})
  \sin \theta_{\ssvc{x}} = \prod_{\ssvc{x}} d\Lx \, \sqrt{1 - \Lx^2}
   \equiv \prod_{\ssvc{x}} D\Lx \; ,
\ee
namely, 
\be
\label{SD4}
  \int \prod_{\ssvc{y}} d\theta_{\ssvc{y}} \, \sin^2 \theta_{\ssvc{y}}
  \left[ 3 \cos \theta_{\ssvc{x}} G - \sin^2 \theta_{\ssvc{x}}
  (G_{\ssvc{x}}^\prime - G S_{\mathrm{eff}, \ssvc{x}}^\prime) \right]
  \exp(-S_{\mathrm{eff}}) = 0 \; .
\ee
A more compact form for these relations is achieved in terms of total
derivatives,  
\bea
\label{SD_RED}
  0 &=&  \int \prod_{\ssvc{y} \ne \ssvc{x}} d\theta_{\ssvc{y}} \sin^2
  \theta_{\ssvc{y}} \int 
  d\theta_{\ssvc{x}} \, \frac{\delta}{\delta \theta_{\ssvc{x}}} \left\{ 
  \sin^3 \theta_{\ssvc{x}} \, G \, \exp(-S_{\mathrm{eff}}) \right\} \nn \\
  &=& \int \prod_{\ssvc{y} \ne \ssvc{x}} d\theta_{\ssvc{y}} \sin^2
  \theta_{\ssvc{y}} \int 
  d(\cos \theta_{\ssvc{x}})  \,
  \frac{\delta}{\delta (\cos \theta_{\ssvc{x}})} \left\{ 
  \sin^3 \theta_{\ssvc{x}} \, G \, \exp(-S_{\mathrm{eff}}) \right\}
  \; .
\eea
Note that the $\sin^3 \theta$ term  ensures the absence of
surface terms. With (\ref{SD_RED}) we have found the Schwinger--Dyson
relations of the \textit{reduced} theory involving only the invariant
$L = \cos \theta$. We do not have a simple geometrical explanation for the
invariance of the reduced Haar measure $\mathcal{D}L$ leading to
(\ref{SD_RED}). The $SO(4)$ symmetry of the measure $\mathcal{D}\pol{}$,
however, is very natural.  

In terms of the Polyakov loop $\Lx$, (\ref{SD4}) is the expectation
value 
\be
\label{SD5}
  \langle 3 \Lx G - (1 - \Lx^2) (G_{\ssvc{x}}^\prime - G S_{\mathrm{eff},
  \ssvc{x}}^\prime) \rangle = 0 \; .
\ee
Comparing with (\ref{SD2}) we notice that it does not matter whether the
expectation value is taken with the full or reduced Haar measure as long
as $G = G[L]$. If we insert the ansatz (\ref{S_EFF}), the
Schwinger--Dyson equations (\ref{SD5}) become a linear system for the
couplings $\lambda_a$,
\be
\label{SD6} 
  \sum_a \langle (1 - \Lx^2) \, G S_{a, \ssvc{x}}^\prime \rangle \, \lambda_a
  = \langle (1 - \Lx^2) \, G_{\ssvc{x}}^\prime \rangle - 3 \, \langle \Lx G
  \rangle \; .
\ee
To solve this unambiguously we need as many independent operators $G$
as there are couplings $\lambda_a$. A particularly natural procedure,
which also turns out to be rather stable numerically, is to choose $G
\equiv S_{b, \ssvc{y}}^\prime$. Any of these operators contains an odd number of
$\Lx$'s so that the minimal set of Schwinger--Dyson equations relates
only nontrivial expectation values,
\be
\label{SD7}
  \sum_a \langle (1 - \Lx^2) \, S_{b, \ssvc{y}}^\prime S_{a,
  \ssvc{x}}^\prime \rangle \, \lambda_a 
  = \langle (1 - \Lx^2) \, S_{b, \ssvc{y} \ssvc{x}}^{\pprime} \rangle - 3
  \, \langle \Lx S_{b, \ssvc{y}}^\prime \rangle \; .
\ee
At this stage, keeping $\vc{x}$ and $\vc{y}$ fixed, the problem of
determining the couplings $\lambda_a$ is well posed
mathematically. Numerically, of course, it is better to use all 
the information one can get, for instance by scanning through all
possible distances $x \equiv |\vc{x} - \vc{y}|$, $x <
N_s/2$. The resulting overdetermined system is then solved by 
least--square methods.  Another possibility is to add new equations to
(\ref{SD7}) by choosing further appropriate monomials or
polynomials in $\Lx$ for the operator $G$. This philosophy will be
extensively adopted in Section~6. Before that, however, we will try to
proceed in a (semi--)analytical fashion.

\section{Single--site distributions of Polyakov loops}

\subsection{Definitions} 

From the effective action of Polyakov loops $S_{\mathrm{eff}}[L]$ one
can derive new probability densities by integrating over (part of)
the loop variables $L$. Of course, this amounts to some kind of
course--graining so that via the new densities one will only have
access to gross properties of the effective action. Nevertheless, these
densities, if chosen properly, exactly reproduce certain expectation
values calculated within the full effective ensemble. Consider, for
instance, the local moments,
\be
\label{LPEXP1}
  \ell_p \equiv \bra \Lx^p \ket \equiv Z^{-1} \int \prod_{\ssvc{y}} D\Ly \,
  \Lx^p \exp  \left(- S_{\mathrm{eff}} [L] \right) \; ,
\ee
where, as usual, the partition function $Z$ is the integral over
$\exp(-S_{\mathrm{eff}})$. Splitting off the $\Lx$--integration, 
(\ref{LPEXP1}) can be rewritten as
\be
\label{LPEXP2}
   \ell_p = \bra \Lx^p \ket \equiv \int_{-1}^1 D\Lx \, \Lx^p \, p_{W}
   [\Lx] \equiv  \bra \Lx^p \ket_W \; ,
\ee
with the probability density $p_{W}$ obtained via integrating
over all $\Ly \ne \Lx$,
\be
\label{PW1}
  p_{W} [\Lx] \equiv Z^{-1}  \int \prod_{\ssvc{y} \ne \ssvc{x}} D\Ly
  \,  \exp \left( -S_{\mathrm{eff}} [\Ly] \right) \equiv
  Z^{-1} \exp \left( - W [\Lx] \right) \; .
\ee
Due to translational invariance, $p_{W}$ (like $\ell_p$) does
not depend on the site $\vc{x}$. Thus, $DL \, p_{W}[L]$ is the
probability to find the value of the Polyakov loop in the interval $[L ,
L + dL]$. The $\mathbb{Z}_2$--symmetry of the effective action implies
that the power $p$ in (\ref{LPEXP1}) and (\ref{LPEXP2}) has to be even,
$p = 2q$, at least for finite volume (no spontaneous symmetry
breaking). Therefore, knowing $p_W$ gives access to all local moments 
$\ell_{2q}$ and (by taking the logarithm) to all local cumulants
$c_{2q}$ as well. A particularly important quantity is the Binder cumulant
\cite{binder:81a,fingberg:93}, defined as the quotient
\be
\label{BINDER}
  b_4  \equiv \frac{c_4}{c_2^2} = \frac{\ell_4}{\ell_2^2} - 3 \; ,
\ee
which measures the deviation from a Gaussian distribution. This 
will be analysed in some detail later on.

From the definition (\ref{PW1}) it is obvious that $p_W$ is blind
against spatial correlations of Polyakov loops. In other words, one
cannot calculate two--point functions like $G_{\ssvc{x}\ssvc{y}} \equiv
\langle \Lx \Ly \rangle$. In principle, this can be remedied by a slight
generalization of (\ref{PW1}). To this end we define a new probability density
depending on $\Lx$ \textit{and} $\Ly$,
\be
  p_{W_2} [\Lx , \Ly] \equiv Z^{-1} \int \prod_{\ssvc{z} \ne \ssvc{x},
  \ssvc{y}} D\Lz \, \exp \left( - S_{\mathrm{eff}} [L]
  \right) \equiv  Z^{-1} \exp \left(- W_2 [\Lx , \Ly] \right) \; .
\ee
Then, one can calculate the following two--point correlators,
\be
  \bra \Lx^p \Ly^q \ket = \int D\Lx \, D\Ly \,  \Lx^p \Ly^q  \, p_{W_2}
  [\Lx , \Ly] \; .
\ee
Obviously, $p_W$ and $p_{W_2}$ are related according to
\be
  p_W [\Lx] = \int D\Ly \, p_{W_2} [\Lx , \Ly] \; .
\ee
If there were no correlations, one would have factorization, $p_{W_2}
[\Lx , \Ly] = p_W [\Lx] p_W [\Ly]$.

\subsection{Determination of single--site distributions}

At first glance, there seems to be not much of a gain by introducing 
densities like the single--site distribution $p_{W}$. Note, however,
that $p_W [L]$ is much simpler than our original density $p_S \equiv Z^{-1}
\exp(-S_{\mathrm{eff}})$ which depends on $N_s^3$ variables rather than
just one. In addition, $p_W$ can be obtained  rather easily from our
Monte Carlo data. The results are fairly smooth 
histograms which are displayed in Figure~\ref{FIG:PW} (for details see
App.~\ref{APP:HISTO}). The most important observation, however, is the
finding that $p_W$ is \textit{flat} below $T_c$, that is, one has an
\textit{equipartition} for $\Lx$. Apparently, this is a remnant of the
$SO(4)$ symmetry discussed in Section~2. Taking the (negative) logarithm of
$p_W$ we obtain the single--site potential $W[L]$ shown in 
Figure~\ref{FIG:W>}.  
\FIGURE[Ht]{\includegraphics[scale=0.8]{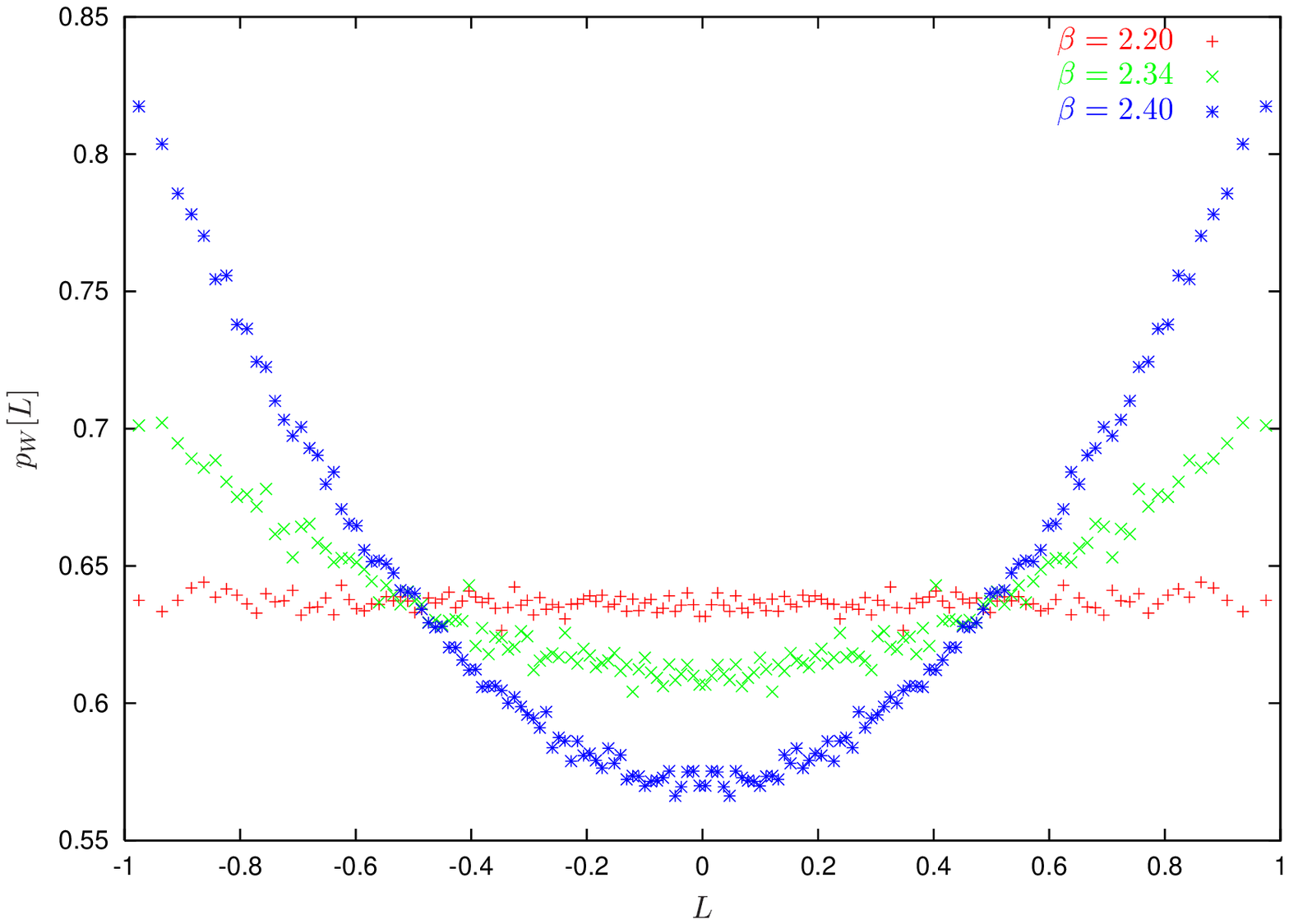}
\caption{Single--site density $p_W [L]$ for temperatures above
($\times$,$*$) and below $T_c$ ($+$). For $T<T_c$ ($\beta < \beta_c \simeq
2.299$), the density is flat, $p_W = 2/\pi$. Input: 200 to 400
configurations, $N_s = 20$.} \label{FIG:PW}}

\FIGURE[Ht]{\includegraphics[scale=0.8]{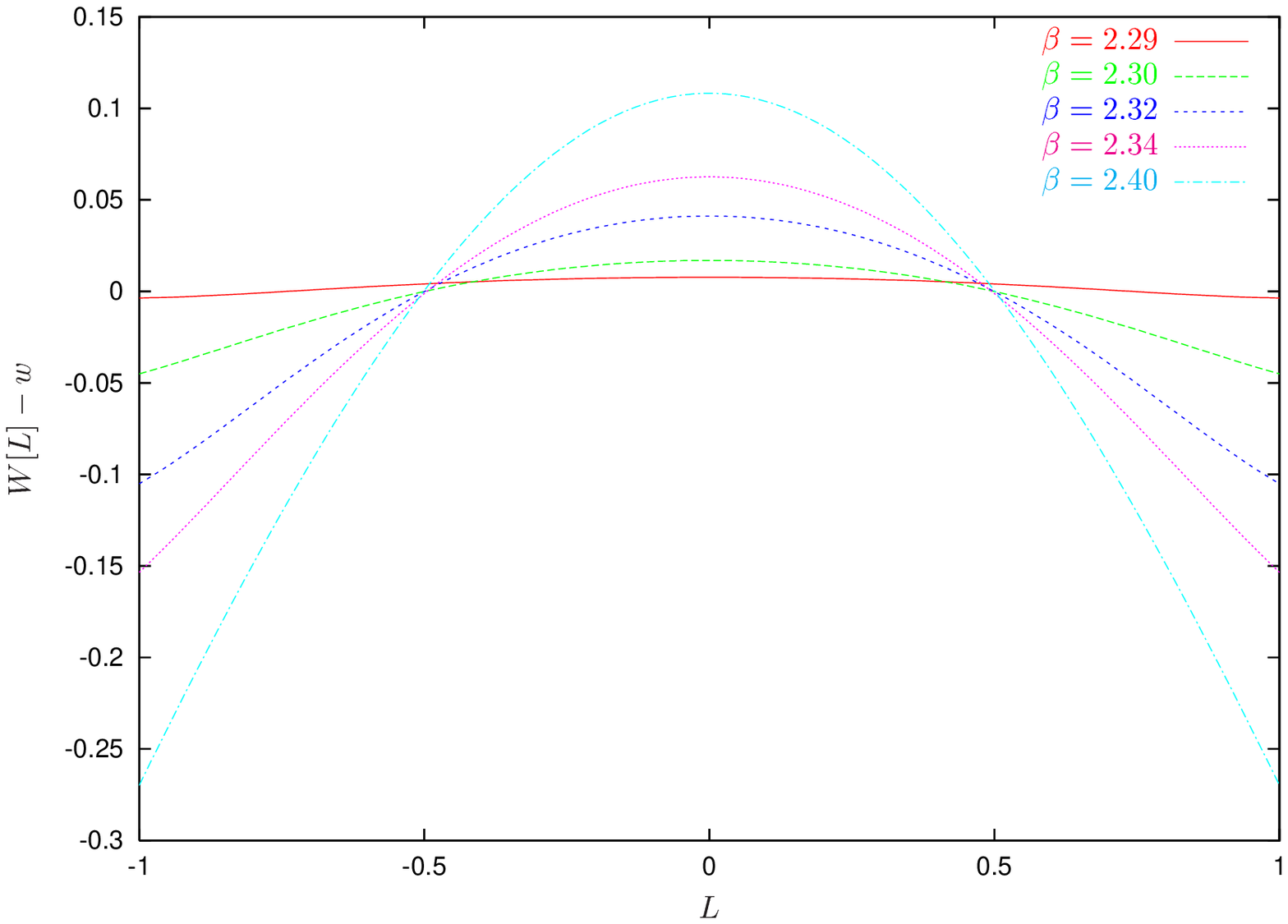}
\caption{The single--site potential $W[L]$ shifted by the constant offset
$w$. For $T < T_c$ ($\beta < \beta_c \simeq 2.299$), $W[L]$ is
flat. Input: 200 to 400 configurations, $N_s = 20$.}
\label{FIG:W>}}

We are thus led to employ the following ansatz for the potential
$W$ from (\ref{PW1}), distinguishing between temperatures  below ($-$)
and above ($+$) the critical value, $T_c$,
\bea
  W_- [L] &=& const \; , \label{W-} \\
  W_+ [L] &=& const' + \sum_k \frac{\kappa_{2k}}{2k} \Lx^{2k} \label{W+} \; .
\eea
Demanding $\bra 1 \ket = 1$ these imply for the density $p_W$,
\bea
  p_W^- [L] &=& \exp(-W_-)/Z_- = 2/\pi \; , \label{PW-} \\
  p_W^+ [L] &=& \exp(- W_{+} [L])/Z_{+} \label{PW+} \; .
\eea
Things are particularly straightforward below $T_c$, so let us discuss
this case first. The result (\ref{PW-}) shows that,  after normalization, the
single--site distribution of Polyakov loops below $T_c$ is known
exactly. Furthermore, it is simple enough so that the associated
(local) moments can be determined analytically, 
\be
\label{L2Q}
  \ell_{2q}^- \equiv \bra L^{2q} \ket_{W_{-}} = \frac{2}{\pi} \int_{-1}^1
  dL \, \sqrt{1 -   L^2} \,   L^{2q} = \frac{1}{\sqrt{\pi}}
  \frac{\Gamma(q +   1/2)}{\Gamma(q+2)} = 2^{-q} \, \frac{(2q -
  1)!!}{(q+1)!} \; . 
\ee
The generating function for these moments can also be calculated
explicitly, 
\be
  Z_- (t) \equiv \bra e^{tL} \ket_{W_-} = \frac{2}{\pi} \int DL \,
  e^{tL}  = \sum_{l \ge 0} \frac{\ell_{2l}^-}{(2l)!} t^{2l} =
  \frac{2}{t} \, I_1 (t) \; , \label{ZT}
\ee
$I_1$ being the standard modified Bessel function. 
For the Binder cumulant (\ref{BINDER}) we thus find the result
\be
\label{BINDER-}
  b_4^- = \frac{\ell_4^-}{(\ell_2^-)^2} - 3 = \frac{1/8}{(1/4)^2} - 3 =
  -1 \;  . 
\ee
We have checked that (\ref{L2Q}) and (\ref{BINDER-}) hold numerically
both for the histograms $p_W$ and the effective Yang--Mills probability
density $p_S$. The results for the Binder cumulant are displayed in
Figure~\ref{FIG:BINDER}. 

\FIGURE[Ht]{\includegraphics[scale=0.7]{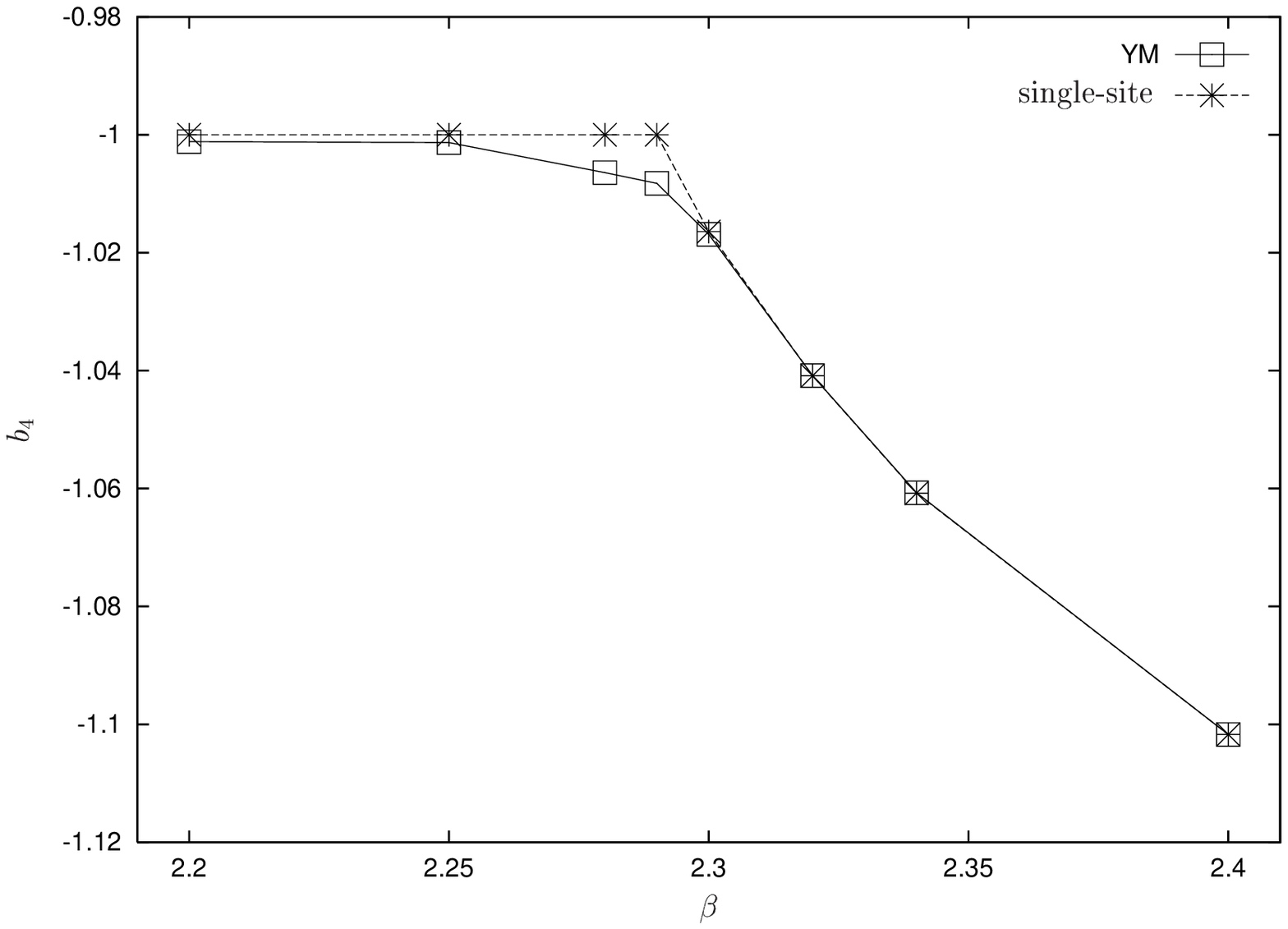}
\caption{The Binder cumulant $b_4$ as obtained from the simulated
Yang--Mills configurations ($\Box$) with $N_s = 20$ compared to the
single--site distribution $p_W$ ($*$). Below $\beta_c \simeq 2.30$, the
exact result (\protect\ref{PW-}) for $p_W$ (i.e.\ $W= const$) has been
used. Above $\beta_c$, $W$ has been fitted to a polynomial (see below).} 
\label{FIG:BINDER}}

It may seem strange that we get a flat distribution $p_W$
below $T_c$. However, this does not imply that the effective potential,
which defines the distribution of the \textit{mean field} $\bar{L}$,
becomes trivial (see Sect.~5). 

To proceed, we have to specify our ansatz for the effective action
beyond (\ref{S_EFF}) and (\ref{S_SVET}). Svetitsky and Yaffe have argued
\cite{svetitsky:82,yaffe:82,svetitsky:86} that, close to the phase
transition, the effective interactions should be short ranged so that
$S_{\mathrm{eff}}$ is of Ginzburg--Landau type,
\be
\label{SVETYAF}
  S_{\mathrm{eff}} = \lambda_0 \sum_{\ssvc{x}, \ssvc{i}} \Lx L_{\ssvc{x}
  + \ssvc{i}} + \sum_{\ssvc{x}} \sum_{k>0} \frac{\lambda_{2k}}{2k}
  (\Lx)^{2k} \equiv \lambda_0 S_0 + \lambda_2 S_2 + \ldots  \; .
\ee
The high--temperature character expansions mentioned in the introduction
yield additional hopping terms of the form $\Lx^p \Ly^q \ldots$
\cite{caselle:95,billo:96,mathur:95}. The relevance of these terms will
be dicussed in Section~6.  

Let us investigate the consequences of the ansatz (\ref{SVETYAF}) for
the single--site distribution. Plugging the former into the definition
(\ref{PW1}) yields
\bea
  e^{-W_-} &=&  \int_{-1}^1 \prod_{\ssvc{y} \ne \ssvc{x}} D\Ly \,
  \exp\left( - \lambda_0 \sum_{\ssvc{y},i} \Ly
  L_{\ssvc{y}   + \ssvc{i}} - \sum_{\ssvc{y} , k>0}
  \frac{\lambda_{2k}}{2k} \Ly^{2k} \right) 
  \nn \\
  &=&  \exp \left( - \sum_{k>0} \frac{\lambda_{2k}}{2k}
  \Lx^{2k} \right) \!
  \int_{-1}^1 \prod_{\ssvc{y} \ne \ssvc{x}} D\Ly \, \exp \left( -
  \lambda_0 \Lx \Mx \right) \exp(- S'_{\mathrm{eff}} [\Ly]) \; , 
   \label{EQUI}
\eea
where, in the second line, we have introduced the field
\be
  \Mx \equiv \pad{S_0}{\Lx}  = \sum_i (L_{\ssvc{x} + \ssvc{i}} +
  L_{\ssvc{x} - \ssvc{i}}) \; ,
\ee
representing the sum of all nearest neighbors of $\Lx$. In addition, we
have defined  a modified action $S'_{\mathrm{eff}}$  which is obtained
from $S_{\mathrm{eff}}$ by setting $\Lx=0$, 
\be
  S'_{\mathrm{eff}}[L] \equiv S_{\mathrm{eff}}[L]\big|_{L_{\ssvc{x}} =
  0}  \; . 
\ee
Now, the left--\-hand side of (\ref{EQUI}) is $2Z_{-}/\pi$ and hence
independent of $\Lx$. Thus we may put $\Lx = 0$ everywhere on the
right--hand side yielding the identity,
\be
\label{NORM}
  2 Z_- /\pi = e^{-W_-} =  \int_{-1}^1 \prod_{\ssvc{y} \ne \ssvc{x}}
  D\Ly \, \exp(- S'_{\mathrm{eff}} [\Ly]) \equiv Z' \; .
\ee
Accordingly, $ e^{-W_-}$ is the partition function associated with action
$S'_{\mathrm{eff}}$. We can go one step further and expand the
exponential containing the nearest--neighbor field $\Mx$ on the
right--hand side of (\ref{EQUI}). This is actually a hopping--parameter
expansion in $\lambda_0$ which, upon using (\ref{NORM}), implies 
\be
\label{HOPPING}
  1 =  \exp \left( - \sum_{k>0} \frac{\lambda_{2k}}{2k}
  \Lx^{2k} \right) \sum_{n \ge 0} \frac{(-\lambda_0)^n}{n!} \Lx^n
  \langle \Mx^n \rangle' \; .
\ee
Here, we have defined  modified expectation values associated with
$S'_{\mathrm{eff}}$ and $Z'$,
\be
\label{MODEXP}
  \bra O[L] \ket' \equiv  \int \prod_{\ssvc{y}\ne \ssvc{x}} D\Ly \,
  O[L] \,  \exp(- S'_{\mathrm{eff}} [L])/Z' \; .
\ee
The $\mathbb{Z}_2$--symmetry of the effective action requires $n$ to be
even, $n = 2m$. Denoting
\be
\label{MUDEF}
  \mu_{2m} \equiv \langle \Mx^{2m} \rangle'   \; ,
\ee
we finally have 
\be
\label{M2M}
  \sum_{m=0}^\infty \frac{\lambda_0^{2m}\mu_{2m}}{(2m)!} \,  \Lx^{2m}  =
  \exp   \left( \sum_{k=1}^\infty  \frac{\lambda_{2k}}{2k} 
  \Lx^{2k} \right) \; .
\ee
To lowest order in $\Lx$ ($m=0$) this consistently reproduces the
normalization (\ref{NORM}), $\bra 1 \ket' = 1 = e^{-W_-}/Z'$. A general
interpretation can be given as follows. To have equipartition requires a
delicate balance between the hopping term ($\lambda_0$) and the
`potential' terms ($\lambda_{2k}$). Setting $\lambda_0 = 0$ (so that the
effective action leads to a product measure) implies that all
$\lambda_{2k}$ have to vanish and \textit{vice versa}: $\lambda_{2k} =
0$ implies $\lambda_0 = 0$. 

To further evaluate the identity (\ref{M2M}) we note that it can be
viewed as a particular example  of a linked--cluster or Mayer
expansion \cite{ursell:27,mayer:37,coester:60} expressing the moments
$\lambda_0^{2m} \mu_{2m}$ in terms of the cumulants 
\be
  \lambda_{2k}' \equiv (2k-1)! \; \lambda_{2k} \; .
\ee
The relation between moments and cumulants can actually be solved for
arbitrary $m$ (see e.g.~\cite{roemer:94}),
\be
\label{CLUSTER}
  \lambda_0^{2m} \mu_{2m} = \sum_{n=1}^m \frac{1}{n!} \sum_{k_1, \ldots
  , k_n = 1 \atop k_1 + \ldots + k_n = m}^m \frac{(2m)!}{(2k_1)! \ldots
  (2k_n)!} \prod_{i=1}^n \lambda_{2k_i}' \; .
\ee
This somewhat clumsy formula yields for the first few orders 
\bea
  \lambda_0^2 \, \mu_2 &=& \lambda_2' \; , \\
  \lambda_0^4 \, \mu_4 &=& \lambda_4' + 3 \, \lambda_2'^2 \; , \\
  \lambda_0^6 \, \mu_6 &=& \lambda_6' + 15 \, \lambda_2' \lambda_4' + 15
  \, \lambda_2'^3 \; , \\
  \lambda_0^8 \, \mu_8 &=& \lambda_8' + 28 \, \lambda_2' \lambda_6' + 35 
  \, \lambda_4'^2 + 210 \, \lambda_4' \lambda_2'^2 + 105 \, \lambda_2'^4
  \; .
\eea
It is quite obvious that by inverting (\ref{CLUSTER}) we can express the
couplings $\lambda_{2k}$ (or cumulants $\lambda_{2k}'$) in terms of the
moments $\mu_{2m}$. Alternatively, one may take the logarithm of
(\ref{M2M}) and compare coefficients. In any case, the first few
cumulants are 
\bea
  \lambda_2' &=& \lambda_0^2 \, \mu_2 \; , \label{LA2} \\
  \lambda_4' &=& \lambda_0^4 \left( \mu_4 - 3 \, \mu_2^2 \right) \; , \\
  \lambda_6' &=& \lambda_0^6 \left( \mu_6 - 15 \, \mu_4 \mu_2 + 30 \, \mu_2^3
  \right) \; , \\
  \lambda_8' &=& \lambda_0^8 \left( \mu_8 - 28 \, \mu_6 \mu_2 + 420 \, \mu_4
  \mu_2^2 - 630 \, \mu_2^4 - 35 \mu_4^2 \right) \; . \label{LA8}
\eea
These identities almost solve our problem of determining
$S_{\mathrm{eff}}$ as they express the unknown couplings $\lambda_{2k}$
in terms of $\lambda_0$ (unknown as yet) and the modified expectation
values $\mu_{2m}$ from (\ref{MUDEF}). 

Things become simple if one allows for only a finite number (say $K$) of
couplings $\lambda_{2k}$ in the Svetitsky--Yaffe action
(\ref{SVETYAF}). Then, there is only a finite number of
independent moments $\mu_{2k}$, $k = 1, \ldots K$. This is quite obvious
from e.g.\ (\ref{LA8}). Setting $\lambda_8 = 0 = \lambda_8'$ determines
the moment $\mu_{8}$ and all higher ones in terms of $\mu_2$, $\mu_4$
and $\mu_6$.  

For $K=1$, (\ref{M2M}) yields the general expression
\be
\label{MU2M1}
  \mu_{2m} = (2m - 1)!! \, \left( \frac{\lambda_2}{\lambda_0^2} \right)^m
  \equiv (2m - 1)!! \; \mu_2^m \; , \quad m = 1, 2, \ldots \; . 
\ee
We thus have found factorization: all higher moments $\mu_{2m}$, $m>1$
can be expressed in terms of the lowest one, $\mu_2 \equiv
\lambda_2/\lambda_0^2$. Of course, this is consistent with
$S_{\mathrm{eff}}$ being quadratic in $\Lx$ (vanishing of quartic and
higher cumulants $\lambda_{2k}'$).

For $K=2$, we have three couplings, $\lambda_0$, $\lambda_2$ and
$\lambda_4$. In this case,  (\ref{M2M}) implies the
following generalization of (\ref{MU2M1}),
\be
\label{MU2M2}
  \mu_{2m} = (2m-1)!! \; \mu_2^m \sum_{k=0}^{[m/2]} {m \choose 2k}
  (2k-1)!! \left( \frac{\mu_4}{3 \mu_2^2} - 1 \right)^k \; ,
\ee
which shows that all moments $\mu_{2m}$ can be expressed in terms of
$\mu_2$ and $\mu_4$. The first two factors in the sum count the number
of ways in which one can form $k$ pairs out of $m$ elements. The term
raised to  power $k$ is actually (one third of) the Binder cumulant
associated with the moments $\mu_{2m}$. If it were zero we would get
back at (\ref{MU2M1}).

Clearly, in order to determine the couplings $\lambda_{2k}$ one does not
want to  calculate the moments $\mu_{2k}$ by performing a new 
and costly Monte Carlo  simulation with the action $S'_{\mathrm{eff}}$,
setting $\Lx = 0$  at a particular site $\vc{x}$. One expects, however,
that, for large lattices, one will have the approximate identity
\be
\label{CRUXID}
  \bra M^{2m} \ket' \simeq \bra M^{2m} \ket \; , \quad m > 0 \; , 
\ee
where the latter expectation is taken in the full Yang--Mills
ensemble. For our numerical evaluation we have tested assumption
(\ref{CRUXID}) as follows. Define the expectation values
\be
  \bra \Mx^{2m} \ket_\Lambda \equiv Z_\Lambda^{-1} \int \prod_{\ssvc{y}}
  D\Ly \, \Mx^{2m}  \exp \big(-S_{\mathrm{eff}} [\Ly] - \Lambda \Lx^2
  \big) \; ,
\ee
so that one has
\be
  \bra \Mx^{2m} \ket = \bra \Mx^{2m} \ket_0 \; , \quad  \bra \Mx^{2m}
  \ket' = \bra \Mx^{2m} \ket_\infty \; .
\ee
If (\ref{CRUXID}) is to hold then $\bra \Mx^{2m} \ket_\Lambda$ must be
approximately independent of $\Lambda$. We have checked this by
simulating  the leading--order action,
\be
  S_\Lambda \equiv  \frac{\lambda_0}{2} \sum_{\bra \ssvc{x} \ssvc{y}
  \ket} \Lx  \Ly + \Lambda \Lx^2 \equiv \lambda_0 \sum_{\ssvc{x},
  \ssvc{i}} \Lx L_{\ssvc{x} + \ssvc{i}} + \Lambda \Lx^2 \; ,
\ee
for different values of $\Lambda$ on a lattice of size $16^3$ with
$\lambda_0 = -0.3$ (symmetric phase). The calculated expectation values
$\bra \Mx^{2} \ket_\Lambda$ displayed in Table~\ref{TABLE:S_LAMBDA-}
show that $\bra \Mx^{2} \ket_\Lambda$ is indeed independent of $\Lambda$
to an accuracy of about 0.5 \%. 

\bigskip

\TABLE[!h]{\parbox{\textwidth}{\centering
\begin{tabular}{|c|cccccc|}\hline
$\Lambda$ & 0 & 1 & 10 & 100 & 1000 & 10000 \\
\hline
$\bra \Mx^{2} \ket_\Lambda$ & 1.951 & 1.947 & 1.962 & 1.954
& 1.939 & 1.961 \\ \hline
\end{tabular}
\caption{The expectation value $\bra \Mx^{2} \ket_\Lambda$ as a function
of the parameter $\Lambda$ suppressing the single--site variable
$\Lx$. Input parameters are $N_s = 16$, $\lambda_0 = -0.3$ (symmetric
phase).}\label{TABLE:S_LAMBDA-}}}   

\medskip

For $T > T_c$, we use the ansatz (\ref{W+}). This implies that formulae
(\ref{HOPPING}--\ref{LA8}) still hold, however, with $\lambda_{2k}$ now
replaced by $\lambda_{2k} - \kappa_{2k}$. We have checked that
the identification (\ref{CRUXID}) also holds in the broken phase
(choosing $\lambda_0 = -1$, see Table~\ref{TABLE:S_LAMBDA+}).  

\bigskip

\TABLE[!h]{\parbox{\textwidth}{\centering
\begin{tabular}{|c|cccccc|}\hline
$\Lambda$ & 0 & 1 & 10 & 100 & 1000 & 10000 \\
\hline
$\bra \Mx^{2} \ket_\Lambda$ & 18.79 & 18.87 & 18.83 & 18.78
& 18.80 & 18.78 \\ \hline
\end{tabular}
\caption{The expectation value $\bra \Mx^{2} \ket_\Lambda$ as a function
of the parameter $\Lambda$ suppressing the single--site variable
$\Lx$. Input parameters are $N_s = 16$, $\lambda_0 = -1$ (broken
phase).}\label{TABLE:S_LAMBDA+}}}

\medskip

The couplings $\kappa_{2k}$ can be obtained by fitting $W_+[L]$ (see
Figure~\ref{FIG:W>}) according to (\ref{W+}). The fit values are displayed
in Tables 3 and 4. 

\bigskip

\DOUBLETABLE[!h]{
\begin{tabular}{|c|cc|}\hline
$\beta$ & $\kappa_2 /2$ & $\kappa_4 /4$ \\ \hline
2.40 &  $-0.4468$ &  0.0703 \\
2.34 &  $-0.2712$ &  0.0526 \\
2.32 &  $-0.1772$ &  0.0261 \\
2.30 &  $-0.0717$ &  0.0120 \\ \hline
\end{tabular}}
{\begin{tabular}{|c|ccc|}\hline
$\beta$ & $\kappa_2 /2$ & $\kappa_4 /4$ & $\kappa_6 /6$ \\ \hline
2.40 &  $-0.4531$ &  0.0901   &  $-0.0152$ \\
2.34 &  $-0.2626$ &  0.0249   &  0.0216 \\
2.32 &  $-0.1612$ & $-0.0259$ &  0.0408 \\
2.30 &  $-0.0666$ & $-0.0087$ &  0.0133 \\ \hline
\end{tabular}}{Two--parameter fit to
$W_{+}[L]$.}{Three--parameter fit to 
$W_{+}[L]$.} 

Summarizing we note that we have good analytical and numerical control
of the single--site distribution $p_W$ or, equivalently, the histograms
displayed in Figure~\ref{FIG:PW}. Below $T_c$, the histogram is flat,
$p_{W}^- = const$, above $T_c$, $W^+ \sim \log p_W^+$ is a simple
polynomial in $L^2$ with coefficients given in Tables 3 and 4.


\section{Determination of the effective action}

The calculation of the couplings $\lambda_{2k}$, $k \ge 0$, in the
effective action proceeds in three steps. First we determine the moments
$\mu_{2m}$ from the Polyakov--loop ensemble using the approximate
identity (\ref{CRUXID}). Second, from (\ref{LA2}--\ref{LA8}), we obtain
the couplings $\lambda_{2k} = \lambda'_{2k}/(2k-1)!$, $k>0$, in terms of
the moments $\mu_{2k}$ and $\lambda_0$. Third, we determine
$\lambda_0$.

The first step consists of straightforward numerics based on our Wilson
ensembles obtained for several values of $\beta$ near
$\beta_c$. The results for the $\mu_{2m}$ are displayed  in
Table~\ref{TABLE:MU}. 

\TABLE[!h]{\parbox{\textwidth}{\centering
\begin{tabular}{|c|cccccccc|}\hline
$\beta$ & 2.20 & 2.25 & 2.28 & 2.29 & 2.30 & 2.32 & 2.34 & 2.40 \\
\hline
$\mu_2$ &1.93& 2.086 &2.242 &2.327 &2.466 &2.946 &3.336 &  4.173  \\
$\mu_4$ &10.16 &11.55 &13.07 &13.89 &15.27 & 20.16& 24.22 &  33.60 \\
$\mu_6$ &80.88 &96.06 &113.0 & 121.7 &137.6 & 194.1 &241.5 & 357.6\\
$\mu_8$ &829.3 &1019 &1237 &1341 &1551 &2297 &2922 & 4536 \\
\hline
\end{tabular}
\caption{The moments $\mu_{2m}$ for different values of the Wilson
coupling $\beta$ ($N_s = 20$).}\label{TABLE:MU}}}

\nin
With the moments $\mu_{2m}$ at hand we find the couplings
\be
\label{LAMBDA_ALPHA}
  \lambda_{2k} = \lambda_0^{2k} \alpha_{2k} \; , \quad k > 0 \; ,
\ee
where the $\alpha_{2k}$ can be expressed in terms of the $\mu_{2k}$
according to (\ref{LA2}--\ref{LA8}). The final step consists in the
determination of $\lambda_0$. To this end we make use of the  
Schwinger--Dyson relations (\ref{SD6}) choosing the operators $G \equiv
\Lx^{2l-1}$ which results in
\be
\label{SDE_LOCAL}
  \langle (1 - \Lx^2) \, \Mx \Lx^{2l-1} \rangle \lambda_0 +
  \sum_{k>0} \langle (1 -
  \Lx^2) \, \Lx^{2k+2l-2}\rangle  \lambda_{2k} = (2l-1) \, \langle (1 -
  \Lx^2) \, \Lx^{2l-2} \rangle - 3 \, \langle \Lx^{2l}  
  \rangle .
\ee
For $T < T_c$, where the single--site distribution
is known exactly, the right--hand side of (\ref{SDE_LOCAL}) 
vanishes. This can either be inferred from the analytical result
(\ref{L2Q}) or by noting that the term in question is a total
derivative,  
\be
  (2l-1) \, \langle (1 -
  \Lx^2) \, \Lx^{2l-2} \rangle - 3 \, \langle \Lx^{2l}  
  \rangle = - \frac{2}{\pi} \int_{-1}^1 dL \, \pad{}{L}
  \left[ (1 - L^2)^{3/2} \, L^{2l - 1} \right] = 0 \; .
\ee
Plugging (\ref{LAMBDA_ALPHA}) into (\ref{SDE_LOCAL}) and dividing by
$\lambda_0$ (assumed to be nonzero) yields  a nonlinear equation of
degree $2k-1$ in $\lambda_0$. With the coefficients $\alpha_{2k}$  and
all nonlocal expectation values (correlators) determined numerically, the
coupling $\lambda_0$ can be obtained straightforwardly. As there are
$2k-1$ solutions we take the one which is approximately independent of
the number $K$ of couplings $\lambda_{2k}$. The resulting values of all
couplings (for $K=2$ and $K=3$) are displayed in Tables \ref{TABLE:K2}
and \ref{TABLE:K3}.  

\TABLE[ht]{\parbox{\textwidth}{\centering
\begin{tabular}{|c|rrrr|rrrr|}\hline
$\beta$     & 2.20 & 2.25 & 2.28 & 2.29 & 2.30 & 2.32 & 2.34 & 2.40 \\
\hline
$\lambda_0$ & $-0.438$ & $-0.473$ & $-0.500$ & $-0.509$ & $-0.630$
& $-0.685$ & $-0.697$ & $-0.725$ \\
$\lambda_2 /2$ & 0.186 & 0.233 & 0.280 & 0.301 & 0.453 & 0.603 &
0.675 & 0.873 \\
$\lambda_4 /4$ & $-0.002$ & $-0.003$ & $-0.005$ & $-0.007$ &
$-0.019$ & $-0.053$ & $-0.088$ & $-0.212$ \\ \hline
\end{tabular}
\caption{Numerical values for the couplings $\lambda_0$ and
$\lambda_{2k}/2k$, $k \le K = 2$. The critical Wilson coupling is $\beta_c =
2.299$ ($N_s = 20$).}\label{TABLE:K2}}}

\TABLE[ht]{\parbox{\textwidth}{\centering
\begin{tabular}{|c|rrrr|rrrr|}\hline
$\beta$     & 2.20 & 2.25 & 2.28 & 2.29 & 2.30 & 2.32 & 2.34 & 2.40 \\
\hline
$\lambda_0$ & $-0.438$ & $-0.476$ & $-0.507$ & $-0.510$ & $-0.628$ &
$-0.690$ & $-0.705$ & $-0.760$ \\
$\lambda_2 /2$ & 0.186 & 0.237 & 0.288 & 0.303 & 0.453 & 0.621 &
0.698 & 0.979 \\
$\lambda_4 /4$ & $-0.002$ & $-0.003$ &  $-0.006$ &  $-0.007$ &
$-0.020$ & $-0.057$ & $-0.093$ & $-0.256$ \\
$\lambda_6 /6$ & 0.000039 & 0.00011 & 0.00027 & 0.00037 &  0.0020 &
0.0106 & 0.0244 & 0.116 \\ \hline
\end{tabular}
\caption{Numerical values for the couplings $\lambda_0$ and
$\lambda_{2k}/2k$, $k \le K = 3$. The critical Wilson coupling is $\beta_c =
2.299$ ($N_s = 20$).}\label{TABLE:K3}}}

With the effective couplings determined we are in the position to check
our results by simulating the effective action. For both $\beta = 2.20$
and $\beta = 2.40$ we have produced 10000 configurations
distributed according to $S_{\mathrm{eff}}$ using the couplings from
Table~\ref{TABLE:K3}.

\FIGURE[!b]{\includegraphics[scale=0.7]{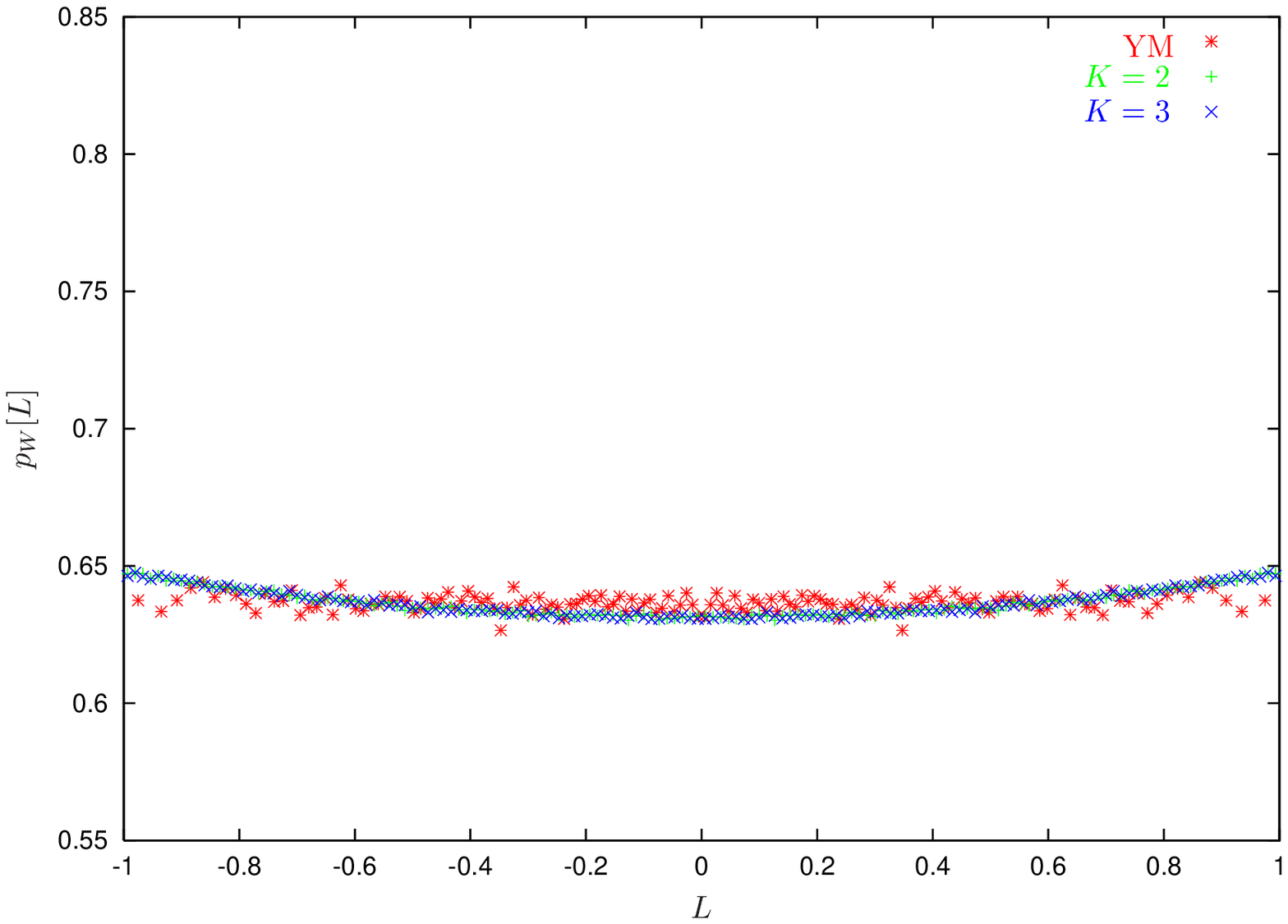}
\caption{Comparison of single--site histograms based on simulating
Yang--Mills ($*$) vs. the effective action for $T < T_c$. The curves for
two and three couplings $\lambda_{2k}$, i.e.\ $K=2$ ($+$)  and $K=3$
($\times$), respectively, fall on top of each other. Input: $\beta =
2.20$, $N_s = 20$.} 
\label{FIG:PW-_CHECK}}

\clearpage

\FIGURE[!ht]
{\includegraphics[scale=0.7]{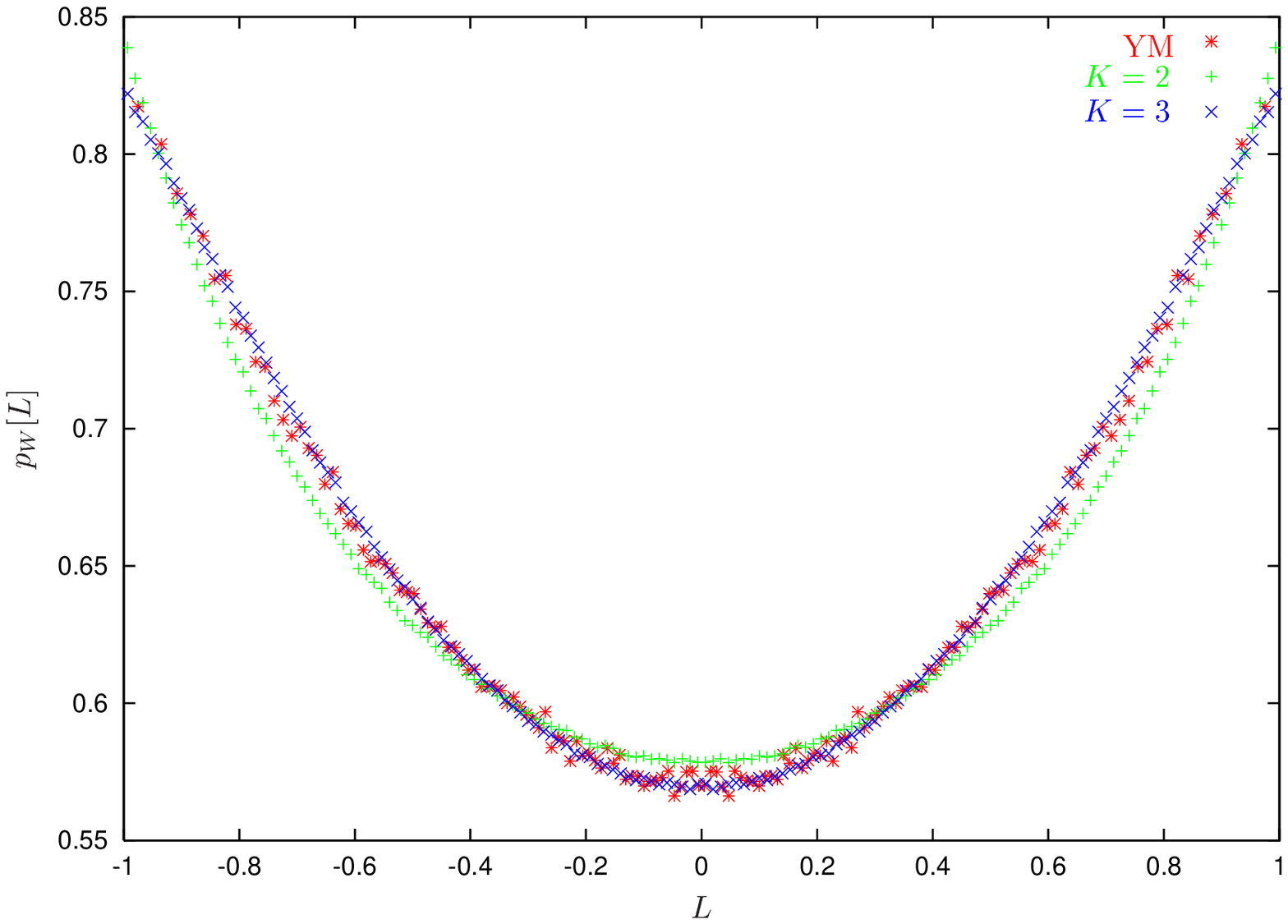}
\caption{Comparison of single--site histograms based on simulating
Yang--Mills ($*$) vs. the effective action with two ($+$) and three ($\times$)
couplings $\lambda_{2k}$ for $T > T_c$. Input: $\beta = 2.40$, $N_s = 20$.}
\label{FIG:PW+_CHECK}}  

In Figures \ref{FIG:PW-_CHECK} and \ref{FIG:PW+_CHECK} we compare the
single--site distributions obtained from 
the effective theory with those of Yang--Mills. The outcome is quite
satisfactory. In particular, one notes that the inclusion of a
$L^6$--term ($K=3$) still improves the matching of the histograms compared to
the case $K=2$. 

A further important check is provided by reproducing the input couplings
of Table~\ref{TABLE:K3} via our IMC
procedure. The results displayed in Table \ref{TABLE:IMC_HISTO} show
quite convincingly that the method works.  If we allow for additional
operators in the numerics (which are not present in the effective
action) the numbers of Table \ref{TABLE:IMC_HISTO} remain unchanged
while the couplings of the new operators are consistently of order
$10^{-5}$, i.e.~compatible with zero. 

\TABLE[ht]{\parbox{\textwidth}{\centering
\begin{tabular}{|c|l|cccc|}\hline
$\beta$     & & $\lambda_0$ & $\lambda_2 $ & $\lambda_4$ & $\lambda_6$ \\
\hline
2.20 & input  & $-0.43803$ & $0.37182$ & $-0.00681$ & $0.00024$ \\
2.20 & output & $-0.43824$ & $0.37351$ & $-0.00621$ & $0.00020$ \\ \hline
2.40 & input  & $-0.76000$ & $1.9572 $ & $-1.0216 $ & $0.69761$ \\
2.40 & output & $-0.76027$ & $1.9605 $ & $-1.0222 $ & $0.69039$ \\ \hline
\end{tabular}
\caption{Comparison of couplings used as input of simulation with
couplings obtained as output of IMC applied to the
effective action.}\label{TABLE:IMC_HISTO}}}

\clearpage

\section{The constraint effective potential}

With an effective action being found, one could go on and calculate the
constraint effective potential \cite{o'raifeartaigh:86} which defines
the distribution of the constant \textit{mean} field, 
\be
  \bar{L} \equiv \frac{1}{\Omega} \sum_{\ssvc{x}} \Lx \; , \quad \Omega =
  N_s^3 \; . \label{MF}
\ee
In perturbation theory, the effective potential has been evaluated long
ago \cite{gross:81,weiss:81}. It describes a `gas' of gluons at high
temperature, i.e.~deep in the deconfined phase. Recent models
for the effective potential which also describe the confined phase are
based on the eigenvalues of the Polyakov loop $\pol{x}$
\cite{meisinger:02a} and not just their sum $\Lx$. As stated in the
introduction, this difference becomes obsolete for $SU(2)$.  

It thus seems of interest to investigate the effective potential on the
lattice. This apparently requires further Monte--Carlo simulations of
the effective action $S_{\mathrm{eff}}[L]$ with the mean field $\bar{L}$
held fixed, following the approach adopted in
\cite{o'raifeartaigh:86,fujimoto:88}. It
turns out, however, that these additional efforts can be avoided by
making use of some statistical properties of the single--site 
distribution $p_W$ discussed in Section~3. 

The constraint effective potential $V$ is defined in terms of the
probability density of the mean field (\ref{MF}), 
\be
\label{V}
  p_V [\bar{L}] \equiv Z_V^{-1} e^{- \Omega V[\bar{L}]} \equiv Z^{-1}
  \int  \mathcal{D}L \, 
  \delta\Big(\bar{L} - \Omega^{-1} \sum_{\ssvc{x}} \Lx \Big) \exp(-
  S_{\mathrm{eff}}[L]) \; ,
\ee
with the normalization $Z_V$ given by the partition function
\be
  Z_V \equiv Z_V (0) \equiv \int_{-1}^1 d\bar{L} \, e^{- \Omega V[\bar{L}]}
  \; .
\ee
In what follows, we will try to obtain the mean--field distribution
$p_V$ from the single--site distribution $p_W$. We note, first of all,
that, due to translational invariance, the first moments coincide, 
\be
  \bra \bar{L} \ket_V \equiv \int d\bar{L} \, \bar{L} \, p_V [\bar{L}]
  = \Omega^{-1} \sum_{\ssvc{x}} Z^{-1} \int \prod_{\ssvc{y}} D\Ly \,  \Lx \,
  e^{-S_{\mathrm{eff}}[L]} \equiv \bra L \ket_W  \equiv \bra L \ket 
  \; . 
\ee
The higher moments, on the other hand, are different,
\bea
  \bra L^p \ket_W &=& \int \prod_{\ssvc{y}} D\Ly \,  \Lx^p \,
  e^{-S_{\mathrm{eff}}[L]} = \bra L^p 
  \ket \label{LP} \\
  \bra \bar{L}^p \ket_V &=& \Omega^{-p} \sum_{\ssvc{x}_1 , \ldots
  \ssvc{x}_p} \bra L_{\ssvc{x}_1} \ldots L_{\ssvc{x}_p} \ket \equiv
  \chi^{(p)} \; .
\eea
For the mean--field distribution we thus get generalized susceptibilities
$\chi^{(p)}$, while $p_W$ yields expectation values of arbitrary powers
of $L$ at a single spatial site, taken in the ensemble of Polyakov loops
extracted from Yang--Mills. This has been discussed at length in Section~3. 

To obtain a connection between arbitrary moments we suppose that the
generating functions associated with $p_V$ and $p_W$ are related
according to 
\be
  Z_V (t) \equiv \bra \exp{t\bar{L}} \ket_V = \bra \prod_{\ssvc{x}} \exp
  (t\Lx /\Omega)\ket_V \stackrel{!}{\simeq} \prod_{\ssvc{x}} \bra
  \exp(tL/\Omega)  \ket_W \equiv [Z_W (t/\Omega)]^\Omega  \; .
\ee
Here, we have made the assumption that only a small fraction of the
random variables $\{ \Lx: \vc{x} \in \Omega \}$ are statistically
dependent. This is justified for large volumes and short--range
correlations. According to the law of large numbers we expect the
collective random variable $\bar{L} = \sum_{\ssvc{x}} \Lx /\Omega$ to have
a Gaussian distribution if the $\Lx$ are randomly
distributed\footnote{Note, however, that with $\bar{L}$ being a compact
variable, we cannot expect a Gaussian in a strict mathematical
sense.}. Let us check to which extent this is realized.  

Below $T_c$, $Z_W \equiv Z_-$ is exactly known from (\ref{ZT}) so that
\be
  Z_V (t) \simeq \left[ \frac{2\Omega}{t} I_1 (t/\Omega) \right]^\Omega =
  \sum_k \frac{t^{2k}}{(2k)!} \bra \bar{L}^{2k} \ket \; , \quad Z_V (0)
  = 1 \; .
\ee
Thus, by expanding the Bessel function (to power $\Omega$) we know all
moments or susceptibilities of $p_V$. Explicitly, one finds
\bea
  \bra \bar{L}^2 \ket_V &=& \frac{1}{4\Omega} \; , \label{CHI2}\\
  \bra \bar{L}^4 \ket_V &=&  \frac{1}{8\Omega^3} + \frac{3(\Omega
  -1)}{16\Omega^3}  \; , \label{CHI4}\\
  \bra \bar{L}^6 \ket_V &=&  \frac{5}{64 \Omega^5} + \frac{15
  (\Omega-1)}{32 \Omega^5} + \frac{15(\Omega - 1)(\Omega -2)}{64
  \Omega^5}  \; .
\eea
In the large--volume limit, $\Omega \to \infty$, the leading terms yield
\be
  \bra \bar{L}^{2k} \ket_V = \frac{(2k-1)!!}{(4\Omega)^k} = (2k-1)!!
  \, \bra \bar{L}^2 \ket_V^k \; , \label{LLV}
\ee
an identity typical for a Gaussian distribution. As a cross check, we
calculate the Binder cumulant associated with $p_V$. From (\ref{CHI2})
and (\ref{CHI4}) we have
\be
  b_{4,V} \equiv \frac{\bra \bar{L}^4 \ket_V}{\bra \bar{L}^2 \ket_V^2} -
  3 = -\frac{1}{\Omega} \; , 
\ee
which obviously vanishes in the infinite--volume limit in accordance
with (\ref{LLV}). Summing up the  moments (\ref{LLV}), we obtain the
large--volume partition function
\be
  Z_V (t) \simeq \exp (t^2/8\Omega)  \; ,
\ee
which turns out to be Gaussian in $t$. Substituting $t = iu$, we have
\be
  Z_V (iu) = \int d\bar{L} \, \exp(-\Omega V[\bar{L}] +
  iu\bar{L}) \simeq \exp(-u^2/8\Omega) \; .
\ee
To extract the mean--field distribution $p_V = \exp
(-\Omega V)/Z_V$ we take the Fourier transform with respect to $u$ and
find
\be
  p_V [\bar{L}] \simeq \sqrt{2\Omega/\pi} \, \exp(-2 \Omega \bar{L}^2) \; ,
\ee
which is a perfect Gaussian distribution with variance
\be
  \sigma^2 \equiv 1/4\Omega = \bra \bar{L}^2 \ket_V \;
  . \label{VARIANCE}
\ee

The fact that $\bar{L}$ is compact does not really matter as in the
large--volume limit assumed, the Gaussian is sharply localized at
$\bar{L} = 0$. This is indeed seen from Figure~\ref{FIG:GAUSSIAN} which
shows that a Gaussian fit to the distribution of $\bar{L}$, 
\be
  p_{V, \mathrm{fit}} [\bar{L}] = 
  \frac{1}{\sqrt{2\pi}\sigma} \, \exp(- \bar{L}^2 / 2\sigma^2) \; ,
  \label{GAUSS}
\ee
works perfectly well.

\FIGURE[Ht]{\includegraphics[scale=0.7]{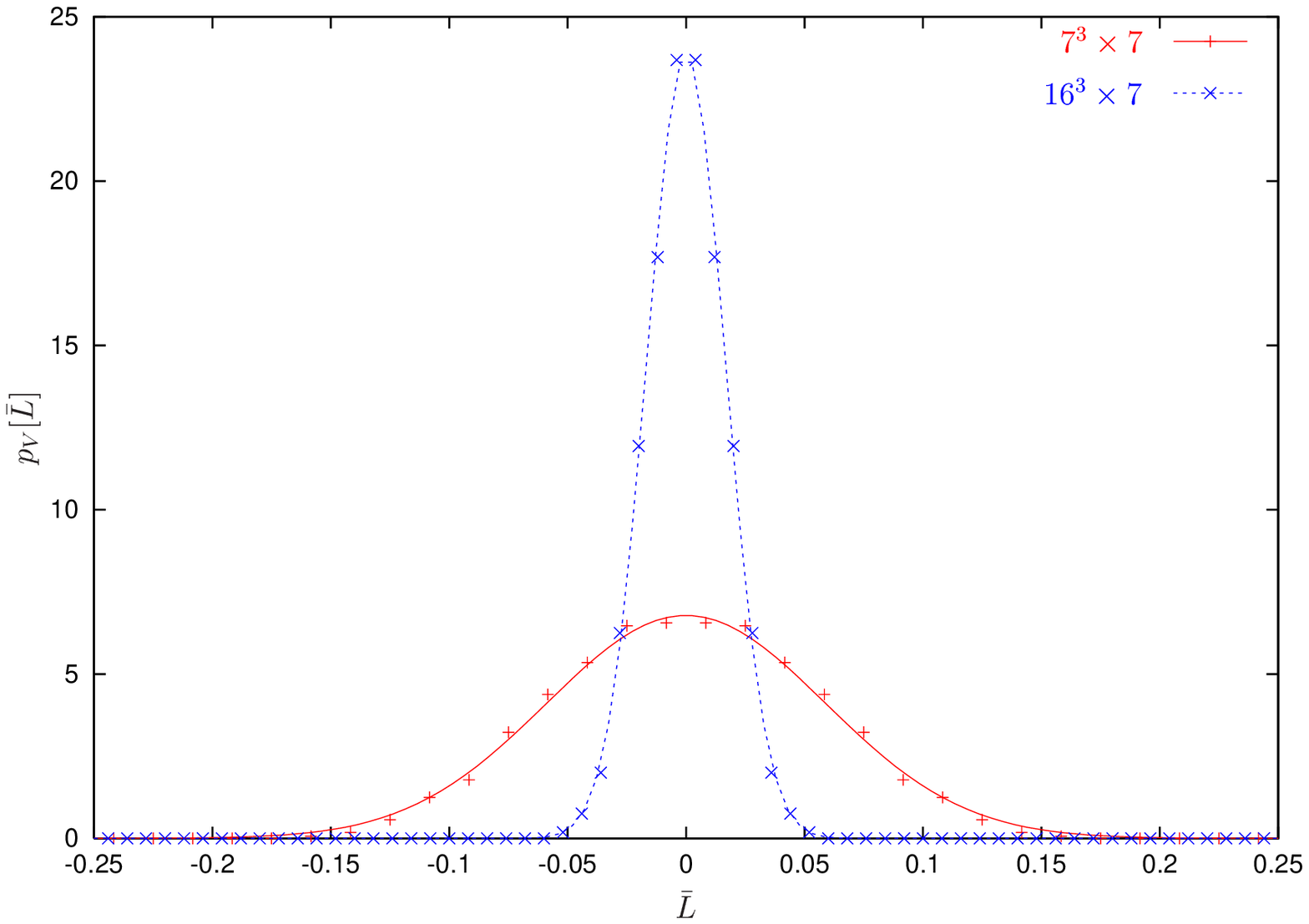}
\caption{Gaussian fits to the distribution $p_V[\bar{L}]$ obtained from
simulating Yang--Mills on lattices of size $16^3 \times 7$ and $7^3
\times 7$. The value $N_t = 7$ for the temporal extension 
corresponds to the symmetric phase.}
\label{FIG:GAUSSIAN}}

This is corroborated by comparing the fit values for $\sigma$ with the
expectation values calculated from Yang--Mills as displayed in
Table~\ref{TABLE:SIGMA} for different volumes and bin sizes.

\TABLE[ht]{\parbox{\textwidth}{\centering
\begin{tabular}{|cccc|}\hline
$\Omega \times N_t$ & config.s/bin & $\sigma$ & $\bra \bar{L}^2
\ket^{1/2}$  \\ \hline
$ 7^3 \times 6 $ & 120 & 0.0837 & 0.0773 \\
$ 7^3 \times 6 $ & 80  & 0.0845 & 0.0773 \\
$ 7^3 \times 7 $ & 120 & 0.0582 & 0.0549 \\
$ 7^3 \times 7 $ & 80  & 0.0588 & 0.0549 \\
$16^3 \times 6 $ & 250 & 0.0249 & 0.0252 \\ 
$16^3 \times 7 $ & 150 & 0.0167 & 0.0164 \\ 
$16^3 \times 7 $ & 250 & 0.0167 & 0.0164 \\ \hline 
\end{tabular}
\caption{Width $\sigma$ of the Gaussian fit (\protect\ref{GAUSS})
compared to the expectation value $\bra \bar{L}^2 \ket^{1/2}$ calculated from
the $SU(2)$ Monte Carlo ensemble. The values for the temporal extension $N_t$
correspond to the symmetric phase.}\label{TABLE:SIGMA}}}

The agreement between the fitted width  and the  
expectation value $\bra \bar{L}^2 \ket^{1/2}$ is quite impressive, in
particular for large volumes, as expected. Due to the approximations
made, however, we do not reproduce the absolute numbers given by
(\ref{VARIANCE}). If we define  
\be
  \gamma \equiv \frac{\sigma^2(\Omega_1)}{\sigma^2(\Omega_2)} =
  \frac{\Omega_2}{\Omega_1}  \; ,
\ee
we get for $\Omega_1 = 7^3$ and $\Omega_2 = 16^3$ the numerical value
$\gamma = (16/7)^3 = 11.94$ while the results of Table~\ref{TABLE:SIGMA}
yield
\bea
  \gamma &=& 11.1 \pm 0.4 \; , \quad N_t = 6 \; , \\
  \gamma &=& 11.5 \pm 0.9 \; , \quad N_t = 7 \; ,
\eea
where the error has been estimated by varying the bin sizes. Thus, at
least for sufficiently low temperature (large $N_t$) we obtain the
correct scaling of the width with the volume. 

\section{Reproducing the two--point function}

The procedure developed so far is based on the single--site distribution
of the Polyakov loop which is under good (semianalytic) control. By
construction, the effective action obtained in this way reproduces the
Yang--Mills distribution quite well (recall Fig.s \ref{FIG:PW-_CHECK}
and \ref{FIG:PW+_CHECK}). At this point it is natural to ask how well
we are reproducing correlators of the Polyakov loop. After all, these
are intimately related to the confining potential ($T < T_c$) or the
Debye mass ($T > T_c$), see e.g.\ \cite{svetitsky:86}. In
Fig.s~\ref{FIG:YMvsL4k_2.2} and \ref{FIG:YMvsL4k_2.4} we compare the
Yang--Mills two--point function with 
the one obtained from the Svetitsky--Yaffe effective action
(\ref{SVETYAF}) using the couplings from Table~\ref{TABLE:IMC_HISTO}.  

\FIGURE[Ht]{\includegraphics[scale=0.85]{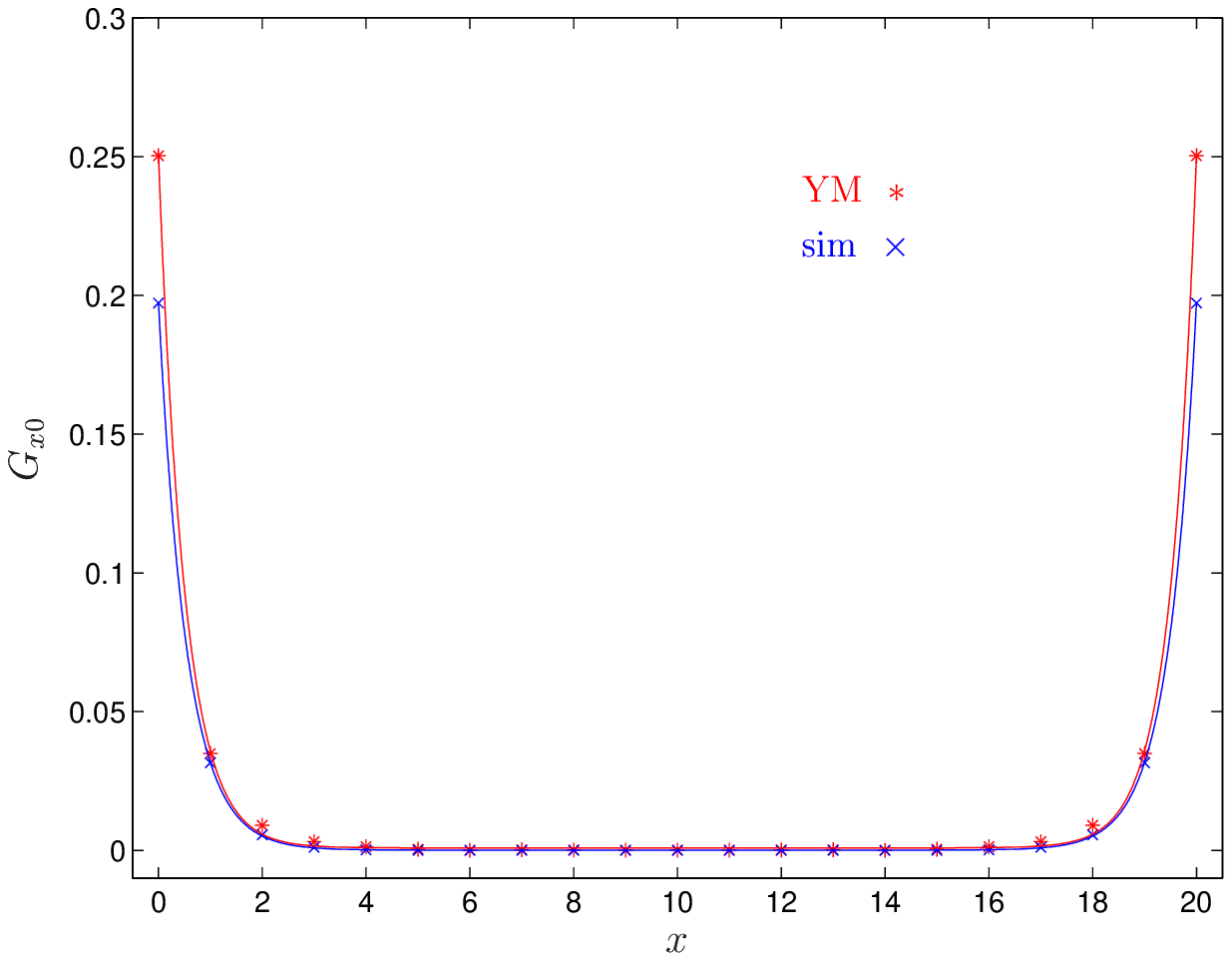}
\caption{The Yang--Mills two--point function (YM) compared to the one
obtained from the Svetitsky--Yaffe effective action with four couplings
(sim). Input: $\beta = 2.20$, $N_s = 20$.}
\label{FIG:YMvsL4k_2.2}}

\FIGURE[Ht]{\includegraphics[scale=0.85]{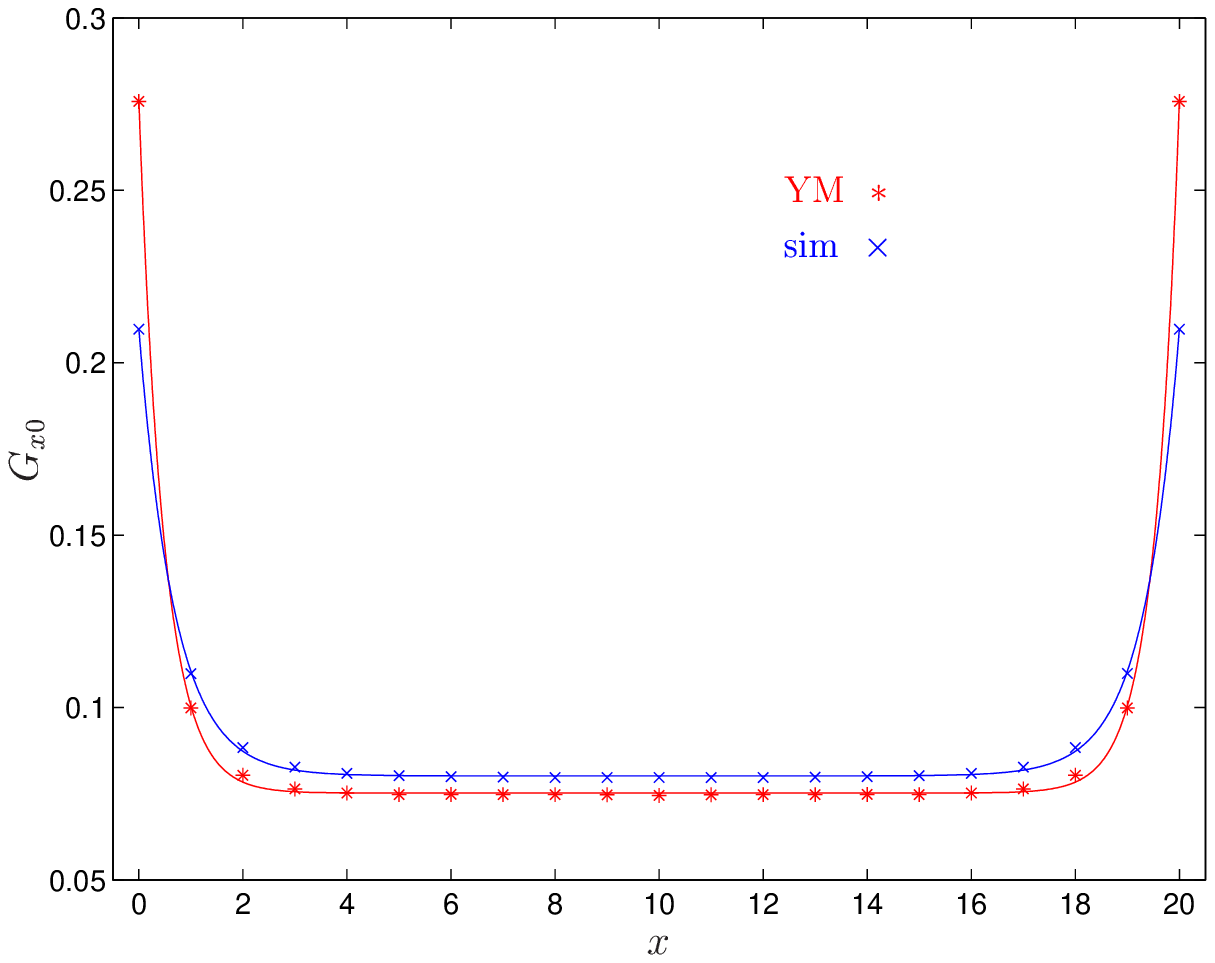}
\caption{The Yang--Mills two--point function (YM) compared to the one
obtained from the Svetitsky--Yaffe effective action with four couplings
(sim). Input: $\beta = 2.40$, $N_s = 20$.}
\label{FIG:YMvsL4k_2.4}}

The figures suggest that we are doing quite well in the symmetric phase
($\beta = 2.20$, i.e.\ $T < T_c$). In the broken phase ($\beta = 2.40$,
i.e.\ $T > T_c$), however, there is room for improvement both in the
exponential decay and the value $\langle L \rangle^2$ of the plateau. To
assess the (dis)agreement quantitatively, we fit all two--point functions
according to
\be
\label{2PTFIT}
  G_{x0} \equiv \langle \Lx L_{\ssvc{0}} \rangle = a \big[ \exp (-bx) + \exp
  \big( -b(N_s - x) \big) \big] + c \; .
\ee
The values for the fit parameters are listed in Table~\ref{TABLE:2PTFIT_4k}
and corroborate the qualitative statements made above.

\TABLE[ht]{\parbox{\textwidth}{\centering
\begin{tabular}{|c|l|ccc|}\hline
$\beta$     & & $a$ & $b $ & $c$  \\
\hline
2.20 & YM  & $0.2493$ & $1.9627$ & $0.0009$  \\
     & sim & $0.1971$ & $1.8309$ & $0.0001$  \\ \hline
2.40 & YM  & $0.2006$ & $2.0715$ & $0.0752$  \\
     & sim & $0.1295$ & $1.4499$ & $0.0802$  \\ \hline
\end{tabular}
\caption{Comparison of the fit parameters from (\protect\ref{2PTFIT})
associated with Fig.s~\protect\ref{FIG:YMvsL4k_2.2} and
\protect\ref{FIG:YMvsL4k_2.4}.}\label{TABLE:2PTFIT_4k}}} 

In order to improve the matching between the effective theory and
Yang--Mills we obviously have to include more operators. In previous
applications of IMC, this has mainly been done for Ising systems
\cite{okawa:88,fortunato:01,fukugita:89,svetitsky:97} or twodimensional
nonlinear sigma models \cite{hasenfratz:94,gottlob:96}. In these cases,
the set of operators is restricted as they square to unity. For the
Polyakov loop, however, the situation is different, as arbitrary (ultralocal)
powers as well as hopping terms associated with arbitrary powers are
allowed, i.e. terms like $L_{\ssvc{x}_1}^{p_1} L_{\ssvc{x}_2}^{p_2}
L_{\ssvc{x}_3}^{p_3} \ldots$. It turns out the the IMC
procedure tends to get destabilized upon including more and more
monomials in $\Lx$. As a result, the values for the couplings depend
rather strongly on the number of operators present \textit{and} of
equations used in the overdetermined linear system. In addition, the
determinants of the matrices to be inverted may become as small as
$10^{-40}$. We thus had to work with symbolic programs like Maple,
setting the number of digits to 60 or even more. Nevertheless, the
instabilities prevailed. Inspired by the
results from the high--temperature expansion on the lattice
\cite{caselle:95,billo:96}, we have tried to overcome these problems by
changing our operator basis from  monomials in $L$ to
\textit{characters}. Being orthogonal class functions, these seem to be
the natural candidates for an economic set of 
operators. At this point it should be noted that for an effective action
with a \textit{finite} number of terms different choices of bases are
\textit{not} equivalent.

As stated in the introduction, for $SU(2)$ the characters can be
expressed as polynomials in the traced Polyakov loop, $L = \tr 
\pol{}/2 = \cos \theta$, according to  
\be
  \chi_j (L) \equiv \frac{\sin \big( (2j + 1) \theta \big)}{\sin \theta}
  = \sum_{p=0}^{[j]} (-1)^p {2j+1 \choose 2p+1} L^{2j-2p} (1 - L^2)^p \;
  , \quad j = 0, \sfrac{1}{2}, 1, \ldots \; .
\ee
This formula allows to reobtain the $L$--representation from the
characters. The first few relations are
\be
  \chi_{1/2} = 2 L \; , \quad \chi_1 = 4 L^2 - 1 \; , \quad \chi_{3/2} =
  8L^3 - 4L  \; , \quad \ldots \; .
\ee
These are sufficient to obtain monomials up to terms of order $\Lx^3
\Ly^3$. To streamline notation it is useful to define a basic link variable
associated with lattice points $\vc{x}$ and $\vc{y}$ and $SU(2)$
`color spin' $j$,
\be
\label{CHARLINK}
  X_{j; \, \ssvc{x}\ssvc{y}} \equiv \chi_j (\Lx) \chi_j (\Ly) \; ,
\ee
which we represent graphically as
\bea
   \Vertex(-30,3){3} \Line(-30,3)(-10,3) \Vertex(-10,3){3} &\equiv& \;
   \charlink{1/2}{x}{y} = 4 \Lx \Ly \; , \label{X1/2}\\
   \Vertex(-30,3){3} \Line(-30,0)(-10,0) \Line(-30,6)(-10,6)
   \Vertex(-10,3){3} &\equiv& \; \charlink{1}{x}{y} \;\;\, = 16 \Lx^2 \Ly^2 - 4
   \Lx^2 - 4 \Ly^2 + 1 \; , \label{X1}\\
   \Vertex(-30,3){3} \Line(-30,0)(-10,0) \Line(-30,6)(-10,6)
   \Line(-30,3)(-10,3) 
   \Vertex(-10,3){3} &\equiv& \;  \charlink{3/2}{x}{y}  = 64 \Lx^3 \Ly^3
   - 32 \Lx \Ly^3 - 32 \Lx^3 \Ly + 16 \Lx \Ly \; 
   , \label{X3/2} \\ 
   &\vdots& \nn 
\eea
A link with $n$ `internal' lines thus corresponds to the representation
labelled by $j = n/2$. These links are the basic building blocks of our
basis of effective operators. The leading order of the high--temperature
expansion \cite{caselle:95,billo:96} is then given by the
nearest--neighbor expression, 
\be
  S_{\mathrm{LO}} \equiv \sum_{\ssvc{x}, \ssvc{i}, j} \lambda_j X_{j;\,
  \ssvc{x}, \, \ssvc{x} + \ssvc{i}} \; ,
\ee
with $\lambda_j$ a known function of the temporal Wilson coupling
$\beta_t$ and extension $N_t$ that decreases rapidly with `color spin'
$j$. If we rewrite the basic link (\ref{CHARLINK}) as $X_{j;\,
\ssvc{x},\, \ssvc{x} + \ssvc{r}}$, we have two parameters controlling
our basis, the representation label $j$ and the effective range (`link length')
$r = |\vc{r}|$. Several test runs of the IMC routines have confirmed
good convergence in $j$ so that we will restrict ourselves to the lowest
representations. The maximum range we allow for is the plaquette
diagonal, i.e.\ $r \le \sqrt{2}$. To further restrict the number of
operators, we limit ourselves to a maximum number of four links of type 
(\ref{CHARLINK}) that can be drawn within a single plaquette. A typical
term, for instance, is thus given by
\be
  \Vertex(-30,-7){3} \Line(-30,-7)(-10,-7) \Vertex(-10,-7){3}
  \Line(-30,-7)(-30,13)  \Vertex(-30,13){3} \Line(-30,-7)(-10,13)
  \Line(-30,13)(-10,-7) \Vertex(-10,13){3} \equiv \;
  X_{1/2; \, \ssvc{x}; \,  \ssvc{x} + \ssvc{i}} \,
  X_{1/2; \, \ssvc{x}, \, \ssvc{x} + \ssvc{j}} \,
  X_{1/2; \, \ssvc{x}, \, \ssvc{x} + \ssvc{i} + \ssvc{j}} \,
  X_{1/2; \, \ssvc{x} + \ssvc{i}, \, \ssvc{x} + \ssvc{j}} \; .
\ee
\TABLE[!ht]{\parbox{\textwidth}{\centering
\begin{tabular}{|ccccccc|} \hline
$$ \Vertex(-5,0){3} \Line(-5,0)(15,0) \Vertex(15,0){3} $$ &
$$ \Vertex(-5,0){3}  \Line(-5,0)(15,20) \Vertex(15,20){3}
\rule[-4mm]{0cm}{1.7cm} $$ & $$ 
\Vertex(-5,0){3} \Line(-5,0)(15,0) \Vertex(15,0){3} \Line(-5,0)(-5,20)
\Vertex(-5,20){3} $$  & $$ \Vertex(-5,0){3} \Line(-5,0)(15,0)
\Vertex(15,0){3} \Line(-5,0)(15,20) \Vertex(15,20){3}   $$ & $$ 
\Vertex(-5,0){3} \Line(-5,0)(15,0) \Vertex(15,0){3} \Line(-5,0)(-5,20)
\Vertex(-5,20){3} \Line(-5,20)(15,0) $$ & $$ \Vertex(-5,0){3}
\Line(-5,0)(15,0) \Vertex(15,0){3} \Vertex(-5,20){3}
\Line(-5,20)(15,20) \Vertex(15,20){3}  $$ & $$ \Vertex(-5,0){3}
\Line(-5,0)(15,0) \Vertex(15,0){3} \Vertex(-5,20){3}
\Line(-5,20)(15,20) \Vertex(15,20){3} \Line(-5,0)(-5,20) $$ \\[.3cm]
$-0.11150$ & $-0.02003$ & $-0.00477$ & $\phantom{-}0.00257$ &
$\phantom{-}0.00368$ & $\phantom{-}0.00191$ & $-0.00052$ \\
$-0.15908$ & $-0.06020$ & $-0.00614$ & $\phantom{-}0.00649$ &
$\phantom{-}0.00535$ & $\phantom{-}0.00547$ & $\phantom{-}0.00003$
\\[1cm] 
$$ \Vertex(-5,0){3}
\Line(-5,0)(15,0) \Vertex(15,0){3} \Vertex(-5,20){3}
\Line(-5,20)(15,20) \Vertex(15,20){3} \Line(-5,0)(15,20) $$ & $$
\Vertex(-5,0){3} \Line(-5,0)(15,0) \Vertex(15,0){3} \Line(-5,0)(-5,20) 
\Vertex(-5,20){3} \Line(-5,0)(15,20) \Vertex(15,20){3}$$ 
& $$ \Vertex(-5,0){3}
\Line(-5,0)(15,0) \Vertex(15,0){3} \Vertex(-5,20){3}
\Line(-5,20)(15,20) \Vertex(15,20){3} \Line(-5,0)(-5,20)
\Line(15,0)(15,20) $$ & $$ \Vertex(-5,0){3}
\Line(-5,0)(15,0) \Vertex(15,0){3} \Vertex(-5,20){3}
\Line(-5,20)(15,20) \Vertex(15,20){3} \Line(-5,0)(-5,20)
\Line(-5,20)(15,0) $$ & $$ \Vertex(-5,0){3} \Line(-5,0)(15,0)
\Vertex(15,0){3} \Line(-5,0)(-5,20) 
\Vertex(-5,20){3} \Line(-5,20)(15,0) \Line(-5,0)(15,20)
\Vertex(15,20){3} $$ & $$ \Vertex(-5,0){3} \Line(-5,-3)(15,-3)
\Line(-5,3)(15,3) \Vertex(15,0){3} $$ & $$ \Vertex(-5,0){3}
\Line(-5,-3)(15,-3) \Line(-5,3)(15,3) \Line(-5,0)(15,0) \Vertex(15,0){3}
$$ \\[.3cm]
$\phantom{-}0.00090$ & $-0.00085$ & $\phantom{-}0.00070$ & $-0.00004$ &
$\phantom{-}0.00021$ & $-0.00833$ & $\phantom{-}0.00008$ \\
$\phantom{-}0.00096$ & $-0.00053$ & $\phantom{-}0.00052$ & $-0.00055$ &
$\phantom{-}0.00001$ & $\phantom{-}0.04305$ & $-0.00061$ \\  
\hline  
\end{tabular}
\caption{Effective operators and couplings for $\beta = 2.20$ (upper
entries) and $\beta = 2.40$ (lower entries), $N_s =
20$.}\label{TABLE:14K_2.2}}}
\nin
Altogether we have 14 operators corresponding to 18 monomials in
$L$. They are displayed in Table~\ref{TABLE:14K_2.2} together with the
couplings associated with them. 
Several comments are in order at this point. By allowing for all possible
distances $x = 0, 1, \ldots, 10$ in the Schwinger--Dyson equations
(\ref{SD7}), we obtain a maximum number of 140 equations for our 14
operators. The values of the couplings remain fairly stable if we vary
the number of equations used in the IMC least--square routine (changes
being approximately 1\% for the relevant couplings). The $\chi^2$ per degree
of freedom is $5 \cdot 10^{-5}$ for the maximum number of 140 equations.

For the operators (\ref{X1/2} -- \ref{X3/2}) we find rapid decrease with
spin $j$. Similarly, if we increase the number of links within the
elementary plaquette, the associated couplings tend to decrease. The
leading order hopping term, $\Vertex(5,3){3} \Line(5,3)(25,3)
\Vertex(25,3){3}$ \hspace{1cm} ($r = 1$) dominates by one order of
magnitude compared to the terms with $r = \sqrt{2}$. This already
indicates that the effective interactions are short--ranged in
accordance with the Svetitsky--Yaffe conjecture. 

If we enumerate the couplings by $g_1, \ldots g_{14}$ from 
left to right, we may express the new effective action as
\be
  \tilde{S}_{\mathrm{eff}} \equiv \sum_{a=1}^{14} g_a \, \tilde{S}_a \; ,
\ee
Note that, according to (\ref{X1/2} -- \ref{X3/2}), the old LO coupling
$\lambda_0$ is given by a (rapidly convergent) series in $j$,
\be
\label{LAMBDA0J}
  \lambda_0  = 4 \, g_1 + 16 \, g_{14} + \mbox{terms with} \;\, j > 3/2
  \; = \; \left\{ \begin{array}{rcl}
                   -0.445 &\quad \mbox{for} \quad& \beta = 2.20 \\
                   -0.646 &\quad \mbox{for} \quad& \beta = 2.40 \\
            \end{array} \right. \quad .
\ee

\FIGURE[Ht]{\includegraphics[scale=0.85]{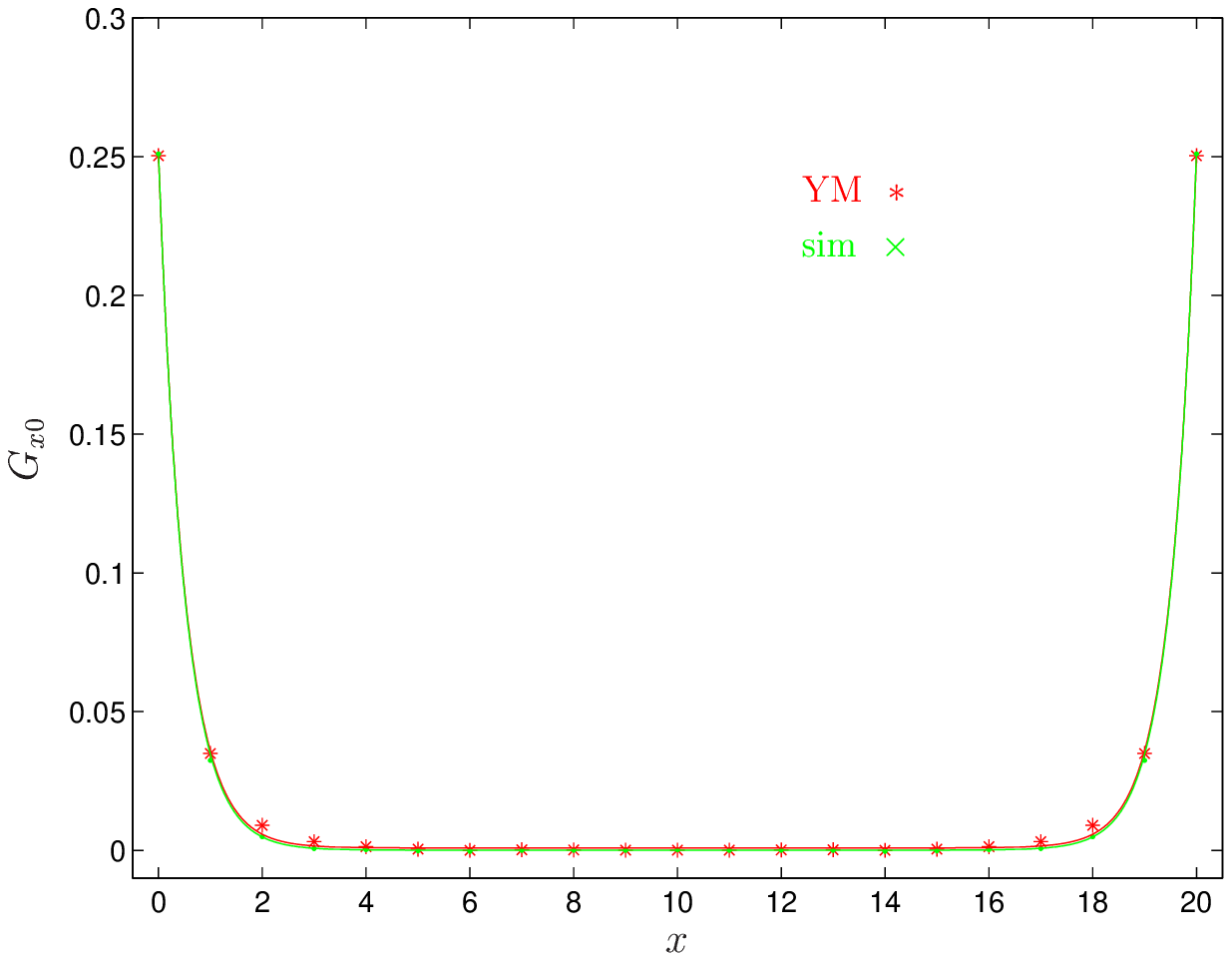}
\caption{The Yang--Mills two--point function (YM) compared to the one
obtained from the character action with 14 couplings
(sim). Input: $\beta = 2.20$, $N_s = 20$.}
\label{FIG:YMvsL14k_2.2}}

\clearpage

These numerical values for $\lambda_0$ agree reasonably well with those of
Table~\ref{TABLE:IMC_HISTO}, where only four operators had been used. 
The benchmark test to be performed, however, is the calculation of the
two--point function $G_{\ssvc{x0}}$ using the new effective couplings
$g_a$. Fig.s \ref{FIG:YMvsL14k_2.2} and \ref{FIG:YMvsL14k_2.4} show that
we have indeed improved the matching between Yang--Mills and the effective
action.  

\FIGURE[Ht]{\includegraphics[scale=0.85]{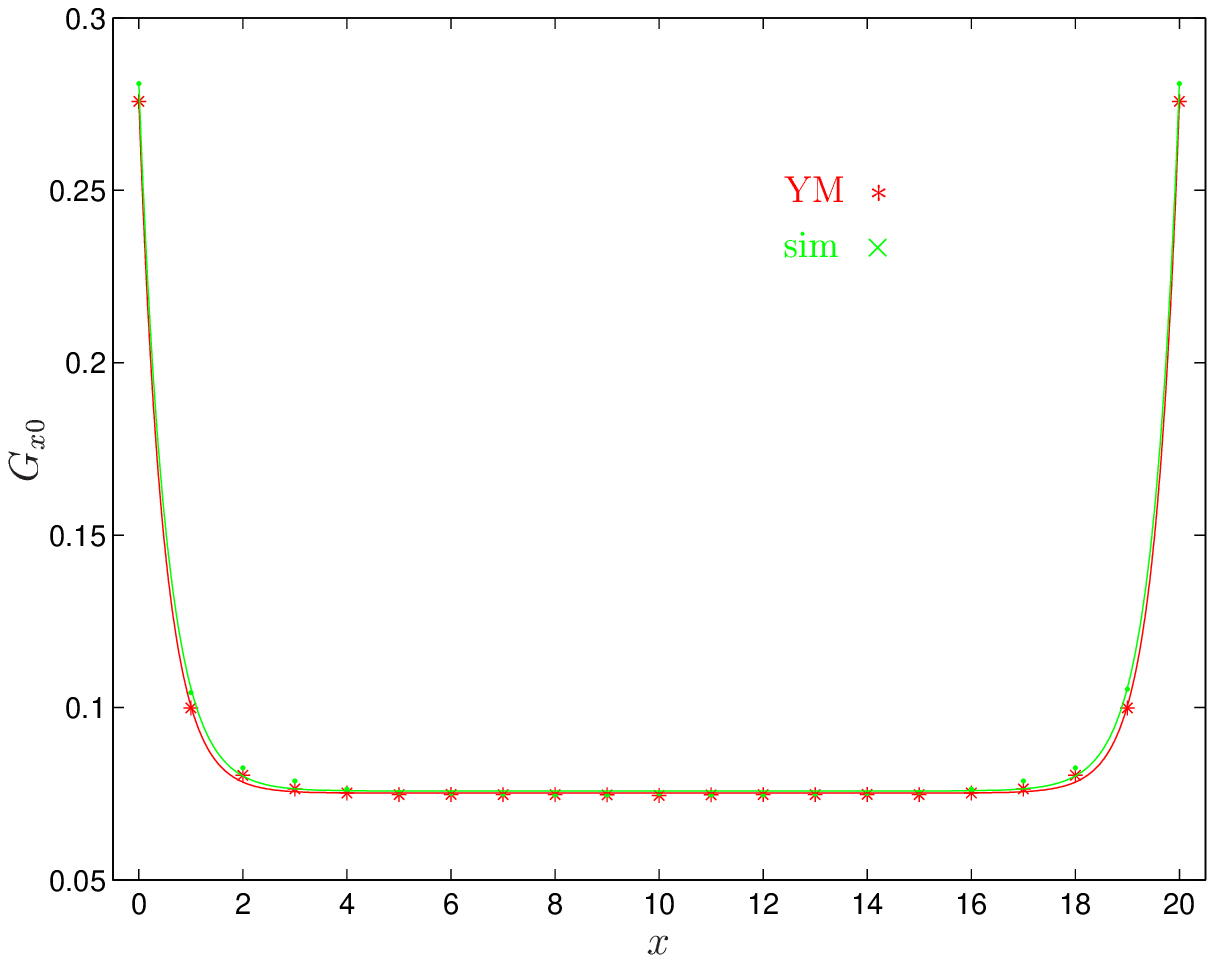}
\caption{The Yang--Mills two--point function (YM) compared to the one
obtained from the character action with 14 couplings
(sim). Input: $\beta = 2.40$, $N_s = 20$.}
\label{FIG:YMvsL14k_2.4}}

This is quantitatively confirmed by repeating the fits of
(\ref{2PTFIT}) and Table~\ref{TABLE:2PTFIT_4k}. The results displayed in
Table~\ref{TABLE:2PTFIT_14k} convincingly  show the improvement in
the effective action, in particular for the broken phase.

\TABLE[!t]{\parbox{\textwidth}{\centering
\begin{tabular}{|c|l|ccc|}\hline
$\beta$     & & $a$ & $b $ & $c$  \\
\hline
2.20 & YM  & $0.2493$ & $1.9627$ & $0.0009$  \\
     & sim & $0.2509$ & $1.9837$ & $0.0001$  \\ \hline
2.40 & YM  & $0.2006$ & $2.0715$ & $0.0752$  \\
     & sim & $0.2051$ & $1.9257$ & $0.0758$  \\ \hline
\end{tabular}
\caption{Comparison of the fit parameters from (\protect\ref{2PTFIT})
associated with Fig.s~\protect\ref{FIG:YMvsL14k_2.2} and
\protect\ref{FIG:YMvsL14k_2.4}.}\label{TABLE:2PTFIT_14k}}} 

\newpage

\section{Summary and discussion}

In this paper we have derived effective actions 
describing the dynamics of the (traced) Polyakov loop variable $\Lx
\equiv \tr \, \pol{x}/2$, and hence of the 
deconfinement phase transition. It has turned out useful, however, to
regard the effective action as being derived from a more general theory
depending on the untraced Polyakov loop $\pol{x}$
\cite{pisarski:00}. This theory is a nonlinear sigma model with target
space $SU(2) \cong S^3$ and hence the symmetry $SU(2) \times SU(2)$
corresponding to left and right multiplication of $\pol{x}$ by group
elements. Although the effective actions in $L$ clearly do not have this
symmetry, it is nevertheless inherited by the functional Haar 
measure which implies novel Schwinger--Dyson equations for
Polyakov loop correlators. In addition, it seems that a remnant of this
symmetry shows up in the single--site distribution $p_W$ of $\Lx$ which is
flat below $T_c$ meaning that $\pol{x}$ is distributed uniformly over the
group manifold. Obviously, it would be desirable to really
\textit{prove} this equipartition for which we have found convincing
numerical evidence. As the single--site distribution of $\Lx$ is exactly
known in the confinement phase, we can give exact predictions for all
moments $\bra L^{2k} \ket$ and for the Binder cumulant, $b_4 = -1$.
Above $T_c$, we have fitted the log--distribution $W \sim \log p_W$ by
polynomials so that also in this case we have good quantitative
control of the distribution.

It turns out that $W[L]$ and a Ginzburg--Landau (or Svetitsky--Yaffe)
effective action $S_{\mathrm{eff}}[L]$ are related in a  manner that is
simple enough to proceed by analytic means. 
Assuming that expectations taken in the effective action are unchanged
if $\Lx$ is changed at a \textit{single} site (another relation valid
numerically but still subject to a proof) we have been able to express
the effective couplings $\lambda_{2k}$ ($k\ne0$) of
$S_{\mathrm{eff}}[L]$  in terms of the
parameters of $W$. The remaining coupling $\lambda_0$ is then determined
by means of the Schwinger--Dyson equations. The single--site
distributions resulting from the effective theory
$S_{\mathrm{eff}}[L]$ agree very well with those obtained directly from
Yang--Mills. Furthermore, the Svetitsky--Yaffe effective action
perfectly fulfills the Schwinger--Dyson equations based on the $SO(4)$
invariance of the Haar measure.  

For the symmetric phase ($T < T_c$) we have also determined the
(constraint) effective potential from the single--site distribution
$p_W$ assuming that the interactions are sufficiently short--ranged such
that  the law of large numbers may be invoked. As expected we obtain a Gaussian
distribution for the mean field $\bar{L}$ if the volume is large and the
temperature small enough.

By definition, one cannot calculate correlations from single--site
distributions. Vice versa, the matching of these distributions does not
imply that the correlation functions match as well. A direct comparison
shows that the two--point functions of the Yang--Mills and
Svetitsky--Yaffe ensembles differ somewhat, in particular in the broken
phase. To improve the matching we have changed our operator basis from
monomials in $\Lx$ to characters, which are orthogonal polynomials in
$\Lx$. Technically, this results in a numerically rather stable inverse
Monte Carlo procedure, even if the number of operators is large. We have
obtained the effective couplings for a total number of 14 operators. The
resulting effective theory has short--range interactions and reproduces
the Yang--Mills two--point function in both phases very well.  

Further research will be devoted to the following issues. The
predictions of the effective actions for the dynamics of the phase
transition should be investigated in detail. This includes an analysis
of the effective potential(s) near and beyond the transition point as well
as calculations of critical exponents. The latter will yield a check whether
the effective action $S_{\mathrm{eff}}[L]$ is indeed in the universality
class of the $\mathbb{Z}_2$--Ising--model. In addition, it should be
possible to generalize the methods developed in this paper to 
higher $SU(N)$ gauge groups. Work in these directions is under way.  

\acknowledgments

The authors thank D.~Antonov, P.~van Baal and J.~Wess for fruitful
discussions and A.~Kirchberg for a careful reading of the
manuscript. The work of T.H.\ was supported by DFG under contract Wi 
777/5-1. 

\appendix

\section{Histograms and bins}
\label{APP:HISTO}

Given a probability density $p_W [L]$ one defines the associated
(cumulative) distribution function
\be
\label{DIST}
  P_W [L] \equiv \int_{-1}^L dL' \, \sqrt{1-L'^2} \; p_W [L'] \; .
\ee
Density $p_W$ and distribution $P_W$ are related to our histograms as
follows. We have a total number $N$ of `events' or `measurements' saying
that a Polyakov loop at site $\vc{x}$ belonging to an arbitrary
configuration takes its value in some prescribed interval
(`bin'). Accordingly, $N$ is a fairly large number,
\be
  N = N_s^3 \times N_{\mathrm{config}} = 20^3 \times 400 = 3.2 \times
  10^6 \; .
\ee
The number of bins (labeled by integers $i$) is denoted by $I$, the
number of events in bin $i$ by $C_i$. This number represents the height
of the $i$th column in the histogram counting the absolute numbers of
events with values in $[L_{i-1} , L_i]$ . The relative counting
\textit{rate} is obtained by normalization,
\be
  \label{CI} c_i \equiv C_i/N = P_W [L_i] - P_W [L_{i-1}] = p_W
  [\bar{L_i}] \, \sqrt{1 - \bar{L}_i^2} \; \Delta L_i \; ,
\ee
where $\Delta L_i \equiv L_i - L_{i-1}$ and $\bar{L_i} \in [L_{i-1}, 
L_i]$ chosen appropriately. The situation is depicted in
Figure~{\ref{FIG:HISTO}}.

\FIGURE[Ht]{\includegraphics[scale=0.5]{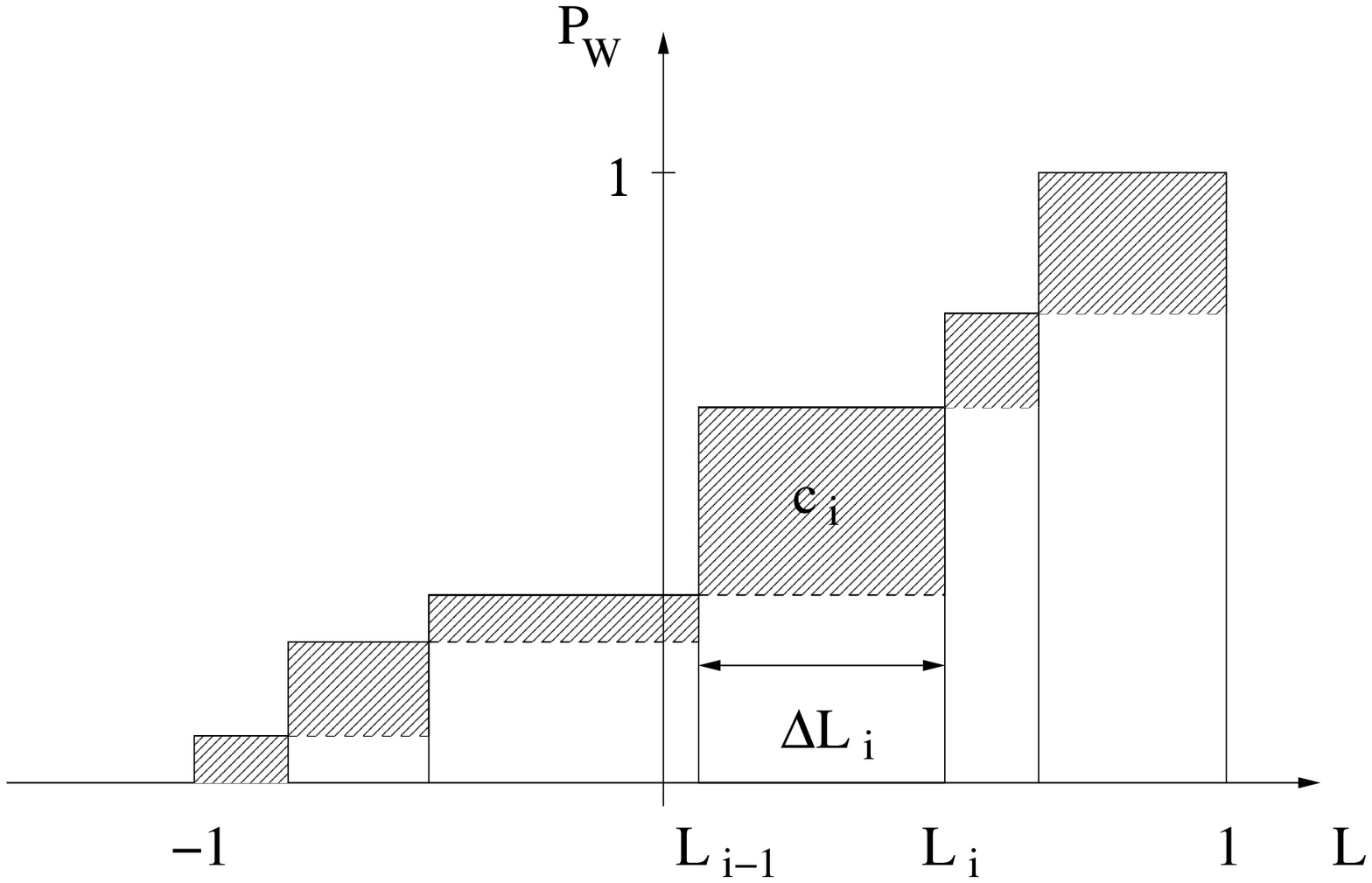}
\caption{General histogram for distribution function $P_W[L]$.}
\label{FIG:HISTO}}
 
Good statistics is achieved if the counting rate $c_i$ is approximately
constant because then all bins will be equally `populated'. This can be
achieved by suitably choosing the bin sizes $\Delta L_i$ which,
however, is somewhat tricky because of the nontrivial measure in
(\ref{DIST}). If we ignore this for the moment and choose an
\textit{equidistant} partition,
\be
  \Delta L_i = \Delta L = 2/I \; ,
\ee
the total count in bin $i$ becomes
\be
\label{CI_EQUI}
  C_i = \frac{2N}{I} \, p_W[\bar{L}_i] \, \sqrt{1 - \bar{L}_i^2} \; .
\ee
This yields rather bad statistics near the boundaries $L = \pm 1$,
in particular for $T > T_c$, due to the suppression by the
measure. For instance, choosing $\beta=2.4$, $I = 100$, i.e. $\Delta L =
1/50$, one typically finds $C_1 \simeq 14000$ data points in the first
bin (near $L=-1$), while the population of the bins near $L=0$ is larger
by a factor of five, $C_{50} \simeq 73000$. The suppression by the
geometry thus `wins' against the density which is peaked near $L = \pm
1$. In the quantity of interest, the probability density,
\be
\label{PW_EQUI}
  p_W [\bar{L_i}] = \frac{C_i}{\sqrt{1 - \bar{L}_i^2}} \, \frac{I}{2N}
  \; ,
\ee
one divides by the measure factor which tends to zero near $L =
\pm 1$. This yields the peaks near $L = \pm 1$ but at the same time
further enhances the statistical error close to the boundaries.  For $T <
T_c$, this is not much of a problem as we have 
equipartition, $p_W [L] = const = p_{W}^- = 2/\pi$, and the density is
known anyhow. For $T > T_c$, however, (\ref{PW_EQUI}) implies that the
bulk of the density is located where the statistical error is
largest. On the other hand, the behavior of $p_W$ in this regime 
determines the higher order couplings $\kappa_{2k}$. The 
lesson to be learned is that the partition should be modified such as to
correctly incorporate the effect of the measure. To this end, we demand
that the counting rate be constant, $c_i = c$, for $T < T_c$, hence,
from (\ref{CI}),  
\be
  c = p_W^- \, \sqrt{1 - \bar{L}_i^2} \; \Delta L_i = 1/I \; .
\ee
Thus, in order to properly take into account the measure, the bin size
$\Delta L_i$ has to be chosen such that  
\be
  \sqrt{1 - \bar{L}_i^2} \; \Delta L_i  = const  = c/p_W^- =
  \frac{1}{I p_W^-} \; .
\ee
This can be achieved by going over to continuum notation,
\be
  c/p_W^- = \int_{L_{i-1}}^{L_i} dL \, \sqrt{1 - L^2} \equiv
  P_W [L_i] - P_W [L_{i-1}] \; ,
\ee
and solving this recursion for $L_i$ numerically with $P_W (L)$ given by
\be
  P_W [L] = \frac{1}{2} \left[L \, \sqrt{1 - L^2} + \arcsin (L) \right]
  \; .
\ee
Alternatively, one may produce an ordered list of all data points for
$L$, and partition this list in such a way that all bins contain the same
number $C$ of `events'. The sampling points $L_i$ are then given by the
smallest (or largest, depending on the counting convention) value of $L$
in bin $i$.   

For $T > T_c$, the density $p_W$ is then given by
\be
  p_W [\bar{L}_i] = \frac{I p_W^-}{N} \, C_i \; .
\ee
This has been displayed in Figure~\ref{FIG:PW}. Obviously, measure effects
are now absent and the difference between 
$C_i$ and $C$ represents the deviation from equipartition.

\bibliographystyle{JHEP.bst}
\bibliography{../../bibfiles/gauge}

\providecommand{\href}[2]{#2}\begingroup\raggedright\begin{thebibliography}{10}

\bibitem{polyakov:78}
A.~M. Polyakov, {\it Thermal properties of gauge fields and quark liberation},
  {\em Phys. Lett.} {\bf B72} (1978) 477.

\bibitem{susskind:79}
L.~Susskind, {\it Lattice models of quark confinement at high temperature},
  {\em Phys.~Rev.} {\bf D20} (1979) 2610.

\bibitem{svetitsky:82}
B.~Svetitsky and L.~Yaffe, {\it Critical behavior at finite-temperature
  confinement transitions},  {\em Nucl.~Phys.} {\bf B210} (1982) 423.

\bibitem{yaffe:82}
L.~G. Yaffe and B.~Svetitsky, {\it First-order phase transition in the SU(3)
  gauge theory at finite temperature},  {\em Phys.~Rev.} {\bf D26} (1982) 963.

\bibitem{caselle:96}
M.~Caselle and M.~Hasenbusch, {\it Deconfinement transition and dimensional
  cross-over in the 3d gauge Ising model},  {\em Nucl. Phys.} {\bf B470} (1996)
  435--453, [\href{http://xxx.lanl.gov/abs/hep-lat/9511015}{{\tt
  hep-lat/9511015}}].

\bibitem{gliozzi:97}
F.~Gliozzi and P.~Provero, {\it The Svetitsky-Yaffe conjecture for the
  plaquette operator},  {\em Phys. Rev.} {\bf D56} (1997) 1131--1134,
  [\href{http://xxx.lanl.gov/abs/hep-lat/9701014}{{\tt hep-lat/9701014}}].

\bibitem{pepe:02}
M.~Pepe and P.~de~Forcrand, {\it Finite size scaling of interface free energies
  in the 3-d Ising model},  {\em Nucl.~Phys.~B (Proc.~Suppl.)} {\bf 106} (2002)
  914, [\href{http://xxx.lanl.gov/abs/hep-lat/0110119}{{\tt hep-lat/0110119}}].

\bibitem{forcrand:03}
P.~de~Forcrand and O.~Jahn, {\it Deconfinement transition in 2+1-dimensional
  SU(4) lattice gauge theory},
  \href{http://xxx.lanl.gov/abs/hep-lat/0309153}{{\tt hep-lat/0309153}}.

\bibitem{polonyi:82}
J.~Polonyi and K.~Szlachanyi, {\it Phase transition from strong coupling
  expansion},  {\em Phys. Lett.} {\bf B110} (1982) 395.

\bibitem{okawa:88}
M.~Okawa, {\it Universality of the deconfining phase transition in
  (3+1)-dimensional SU(2) lattice gauge theory},  {\em Phys. Rev. Lett.} {\bf
  60} (1988) 1805.

\bibitem{fortunato:01}
S.~Fortunato, F.~Karsch, P.~Petreczky, and H.~Satz, {\it Effective Z(2) spin
  models of deconfinement and percolation in SU(2) gauge theory},  {\em Phys.
  Lett.} {\bf B503} (2001) 321,
  [\href{http://xxx.lanl.gov/abs/hep-lat/0011084}{{\tt hep-lat/0011084}}].

\bibitem{banks:83}
T.~Banks and A.~Ukawa, {\it Deconfining and chiral phase transition in quantum
  chromodynamics at finite temperature},  {\em Nucl.~Phys.} {\bf B225} (1983)
  145.

\bibitem{ogilvie:84}
M.~Ogilvie, {\it Effective-spin model for finite-temperature QCD},  {\em
  Phys.~Rev.~Lett.} {\bf 52} (1984) 1369.

\bibitem{svetitsky:86}
B.~Svetitsky, {\it Symmetry aspects of finite-temperature phase transitions},
  {\em Phys.~Rept.} {\bf 132} (1986) 1.

\bibitem{caselle:95}
M.~Caselle, {\it Recent results in high-temperature lattice gauge theories},
  \href{http://xxx.lanl.gov/abs/http://arXiv.org/abs/hep-lat/9601009}{{\tt
  http://arXiv.org/abs/hep-lat/9601009}}. in: \textit{Selected Topics in
  Nonperturbative QCD}, A.~Di Giacomo and D.~Diakonov, eds., Proceedings
  International School of Physics ``Enrico Fermi'', Course CXXX, Varenna,
  Italy, 1995, IOS Press, Amsterdam, 1996.

\bibitem{billo:96}
M.~Billo, M.~Caselle, A.~D'Adda, and S.~Panzeri, {\it Toward an analytic
  determination of the deconfinement temperature in SU(2) l.g.t},  {\em Nucl.
  Phys.} {\bf B472} (1996) 163,
  [\href{http://xxx.lanl.gov/abs/http://arXiv.org/abs/hep-lat/9601020}{{\tt
  http://arXiv.org/abs/hep-lat/9601020}}].

\bibitem{pisarski:00}
R.~Pisarski, {\it Quark-gluon plasma as a condensate of $Z(3)$ Wilson lines},
  {\em Phys.~Rev.} {\bf D62} (2000) 111501(R),
  [\href{http://xxx.lanl.gov/abs/hep-ph/0006205}{{\tt hep-ph/0006205}}].

\bibitem{meisinger:02a}
P.~Meisinger, T.~Miller, and M.~Ogilvie, {\it Phenomenological equations of
  state for the quark-gluon plasma},  {\em Phys.~Rev.} {\bf D65} (2002) 034009.

\bibitem{reinhardt:97b}
H.~Reinhardt, {\it Resolution of Gauss' law in Yang-Mills theory by gauge
  invariant projection: Topology and magnetic monopoles},  {\em Nucl. Phys.}
  {\bf B503} (1997) 505, [\href{http://xxx.lanl.gov/abs/hep-th/9702049}{{\tt
  hep-th/9702049}}].

\bibitem{ford:98}
C.~Ford, U.~G. Mitreuter, T.~Tok, A.~Wipf, and J.~M. Pawlowski, {\it Monopoles,
  Polyakov loops and gauge fixing on the torus},  {\em Ann.~Phys.~(N.Y.)} {\bf
  269} (1998) 26, [\href{http://xxx.lanl.gov/abs/hep-th/9802191}{{\tt
  hep-th/9802191}}].

\bibitem{jahn:98}
O.~Jahn and F.~Lenz, {\it Structure and dynamics of monopoles in axial gauge
  QCD},  {\em Phys. Rev.} {\bf D58} (1998) 085006,
  [\href{http://xxx.lanl.gov/abs/hep-th/9803177}{{\tt hep-th/9803177}}].

\bibitem{thooft:81a}
G.~'t~Hooft, {\it Topology of the gauge condition and new confinement phases in
  non-Abelian gauge theories},  {\em Nucl.~Phys.} {\bf B190} (1981) 455.

\bibitem{binder:81a}
K.~Binder, {\it Critical properties from Monte Carlo coarse graining and
  renormalization},  {\em Phys.~Rev.~Lett.} {\bf 47} (1981) 693.

\bibitem{fingberg:93}
J.~Fingberg, U.~M. Heller, and F.~Karsch, {\it Scaling and asymptotic scaling
  in the SU(2) gauge theory},  {\em Nucl. Phys.} {\bf B392} (1993) 493--517,
  [\href{http://xxx.lanl.gov/abs/hep-lat/9208012}{{\tt hep-lat/9208012}}].

\bibitem{mathur:95}
M.~Mathur, {\it Landau Ginzburg model and deconfinement transition for extended
  SU(2) Wilson action},  \href{http://xxx.lanl.gov/abs/hep-lat/9501036}{{\tt
  hep-lat/9501036}}.

\bibitem{ursell:27}
H.~Ursell, {\it The evaluation of Gibbs' phase integral for imperfect gases},
  {\em Proc.~Cambridge Phil.~Soc.} {\bf 23} (1927) 685.

\bibitem{mayer:37}
J.~Mayer, {\it The statistical mechanics of condensing systems. I},  {\em
  J.~Chem.~Phys.} {\bf 5} (1937) 67.

\bibitem{coester:60}
F.~Coester and R.~Haag, {\it Representation of states in a field theory with
  canonical variables},  {\em Phys.~Rev.} {\bf 117} (1960) 1137.

\bibitem{roemer:94}
H.~R{\"o}mer and T.~Filk, {\em Statistische Mechanik}.
\newblock VCH, Weinheim, 1994.
\newblock (in German).

\bibitem{o'raifeartaigh:86}
L.~O'Raifeartaigh, A.~Wipf, and H.~Yoneyama, {\it The constraint effective
  potential},  {\em Nucl. Phys.} {\bf B271} (1986) 653.

\bibitem{gross:81}
D.~Gross, R.~Pisarski, and L.~Yaffe, {\it QCD and instantons at finite
  temperature},  {\em Rev.~Mod.~Phys.} {\bf 53} (1981) 43.

\bibitem{weiss:81}
N.~Weiss, {\it Effective potential for the order parameter of gauge theories at
  finite temperature},  {\em Phys.~Rev.} {\bf D24} (1981) 475.

\bibitem{fujimoto:88}
Y.~Fujimoto, H.~Yoneyama, and A.~Wipf, {\it Symmetry restoration of scalar
  models at finite temperature},  {\em Phys.~Rev.} {\bf D38} (1988) 2625--2634.

\bibitem{fukugita:89}
M.~Fukugita, M.~Okawa, and A.~Ukawa, {\it Order of the deconfining phase
  transition in SU(3) lattice gauge theory},  {\em Phys.~Rev.~Lett.} {\bf 63}
  (1989) 1768.

\bibitem{svetitsky:97}
B.~Svetitsky and N.~Weiss, {\it Ising description of the transition region in
  SU(3) gauge theory at finite temperature},  {\em Phys.~Rev.} {\bf D56} (1997)
  5395.

\bibitem{hasenfratz:94}
P.~Hasenfratz and F.~Niedermayer, {\it Perfect lattice action for
  asymptotically free theories},  {\em Nucl.~Phys.} {\bf B414} (1994) 785,
  [\href{http://xxx.lanl.gov/abs/hep-lat/9308004}{{\tt hep-lat/9308004}}].

\bibitem{gottlob:96}
A.~Gottlob, M.~Hasenbusch, and K.~Pinn, {\it Iterating block spin
  transformations of the O(3) non-linear sigma-model},  {\em Phys. Rev.} {\bf
  D54} (1996) 1736--1747, [\href{http://xxx.lanl.gov/abs/hep-lat/9601014}{{\tt
  hep-lat/9601014}}].

\end{thebibliography}\endgroup

\end{document}